\documentclass[onecolumn,authoryear]{els-mrw} 

\usepackage{amsmath,amssymb,amsfonts,amsthm,makeidx,graphicx}
\usepackage{txfonts}
\usepackage{helvet}
\usepackage{wrapfig}
\usepackage[resetlabels,labeled]{multibib}
\newcites{Further}{Further Information References}

\newcommand{\aap}{A\&A}
\newcommand{\aj}{AJ}
\newcommand{\apj}{ApJ}
\newcommand{\apjl}{ApJL}
\newcommand{\apjs}{ApJS}
\newcommand{\gca}{GeCoA}
\newcommand{\jcap}{JCAP}
\newcommand{\maps}{MAPS}
\newcommand{\mnras}{MNRAS}
\newcommand{\nat}{Nature}
\newcommand{\rmxaa}{RMxAA}
\newcommand{\ssr}{Space Science Reviews}
\newcommand{\physrep}{Phys. Rep.}
\newcommand{\prd}{Phys. Rev. D}

\newcommand{\het}{$^{3}$He}
\newcommand{\hetp}{$^{3}$He$^{+}$}
\newcommand{\hef}{$^{4}$He}
\newcommand{\hefp}{$^{4}$He$^{+}$}
\newcommand{\lis}{$^{7}$Li}
\newcommand{\HI}{H\,\textsc{i}}
\newcommand{\HeI}{He\,\textsc{i}}
\newcommand{\HII}{H\,\textsc{ii}}
\newcommand{\li}{$^{7}$Li}
\newcommand{\Lya}{Ly$\alpha$}

\begin{document}

\chapter{Big Bang Nucleosynthesis}\label{chap1}

\author[1]{Ryan Cooke}%
% \author[2]{Second Author}%
% \author[1,2]{Third Author}%

\address[1]{\orgname{Centre for Extragalactic Astronomy}, \orgdiv{Department of Physics}, \orgaddress{Durham University, South Road, Durham DH1 3LE, UK}}
% \address[2]{\orgname{Name of Institute}, \orgdiv{Division or Department}, \orgaddress{Address of Institute}}

\articletag{Chapter Article tagline: update of previous edition,, reprint..}

\maketitle

\begin{glossary}[Glossary]
\term{Astration} The nuclear burning of an element by a star.\\
\term{Baryon Asymmetry} The slight favouring of baryons over anti-baryons.\\
\term{Deuteron} The nucleus of a deuterium atom, comprising one proton and one neutron, $^{2}$H.\\
\term{Deuteron bottleneck} Deuterons are the first product of nucleosynthesis, but they are easily destroyed by the high energy tail of the photon blackbody distribution. The deuteron bottleneck is overcome when the photon temperature is much lower than the deuteron binding energy. \\
\term{dex} Shorthand for `decimal exponent'. This unit is usually used to describe a difference in a logarithmic quantity.\\
\term{Isotope} Each chemical element has a fixed number of protons in the nucleus. The isotopes of each chemical element have a different number of neutrons. Some isotopes are stable, while others decay into lighter elements.\\
\term{Metal} In the context of this chapter, a metal is an element that was not produced during Big Bang Nucleosynthesis.\\
\term{Nucleus} The neutrons and protons that comprise an atom.\\
\term{Nuclide} A nuclide is characterized by the number of neutrons and protons in the nucleus of an atom.\\
\term{Quasar} A rapidly accreting supermassive black hole. These objects are often very bright, and can be used to study absorbing material along the line-of-sight.\\
\term{Spite Plateau} The constant \lis/H ratio that is observed in warm, extremely metal-poor halo stars, and represents the current best observational determination of the primordial lithium abundance.\\
\term{Triton} The nucleus of a tritium atom, comprising one proton and two neutrons, $^{3}$H.\\

\end{glossary}

\begin{glossary}[Nomenclature]
\begin{tabular}{@{}lp{34pc}@{}}
BBN &Big Bang nucleosynthesis\\
CMB & Cosmic Microwave Background\\
D/H & Primordial number abundance of deuterons relative to protons, $^{2}{\rm H}/^{1}{\rm H}$\\
\Lya &Lyman-$\alpha$ (i.e. the electronic transition between principal quantum number $n=1\,\leftrightarrow\,n=2$)\\
$\eta$ & The baryon-to-photon ratio\\
H\,\textsc{i} & Neutral hydrogen\\
H\,\textsc{ii} & Ionized hydrogen\\
LTE & Local Thermodynamic Equilibrium\\
$N_{\rm eff}$ & Effective number of neutrino species\\
NSE & Nuclear Statistical Equilibrium\\
Ryd & Shorthand for the unit Rydberg ($1~{\rm Ryd}\simeq13.6~{\rm eV}$)\\
SMC & Small Magellanic Cloud\\
WMAP & Wilkinson Microwave Anisotropy Probe\\
$[{\rm X/Y}]$ & This `square bracket notation' refers to the abundance of element X relative to the abundance of element Y, on a logarithmic scale relative to the Sun: ${\rm [X/Y]}={\rm log}_{10}({\rm X/Y})-{\rm log}_{10}({\rm X/Y})_{\odot}$. For example, a value of $0$ represents the solar abundance, while a value of $-1$ represents one-tenth of the solar abundance.\\
$y_{\rm p}$ & Primordial number abundance of $^{4}{\rm He}/{\rm H}$\\
$Y_{\rm p}$ & Primordial mass fraction of $^{4}{\rm He}$\\
% ECHA &European Chemical Agency\\
% EPM &Equilibrium Partitioning Method Equilibrium Partitioning Method Equilibrium Partitioning Method Equilibrium\hfill\break Partitioning Method\\
% ERA &Ecological Risk Assessment\\
% HC &Hazardous Concentration\\
\end{tabular}
\end{glossary}

\begin{abstract}[Abstract]
\begin{figure}
    \centering
    \includegraphics[width=\columnwidth,angle=0]{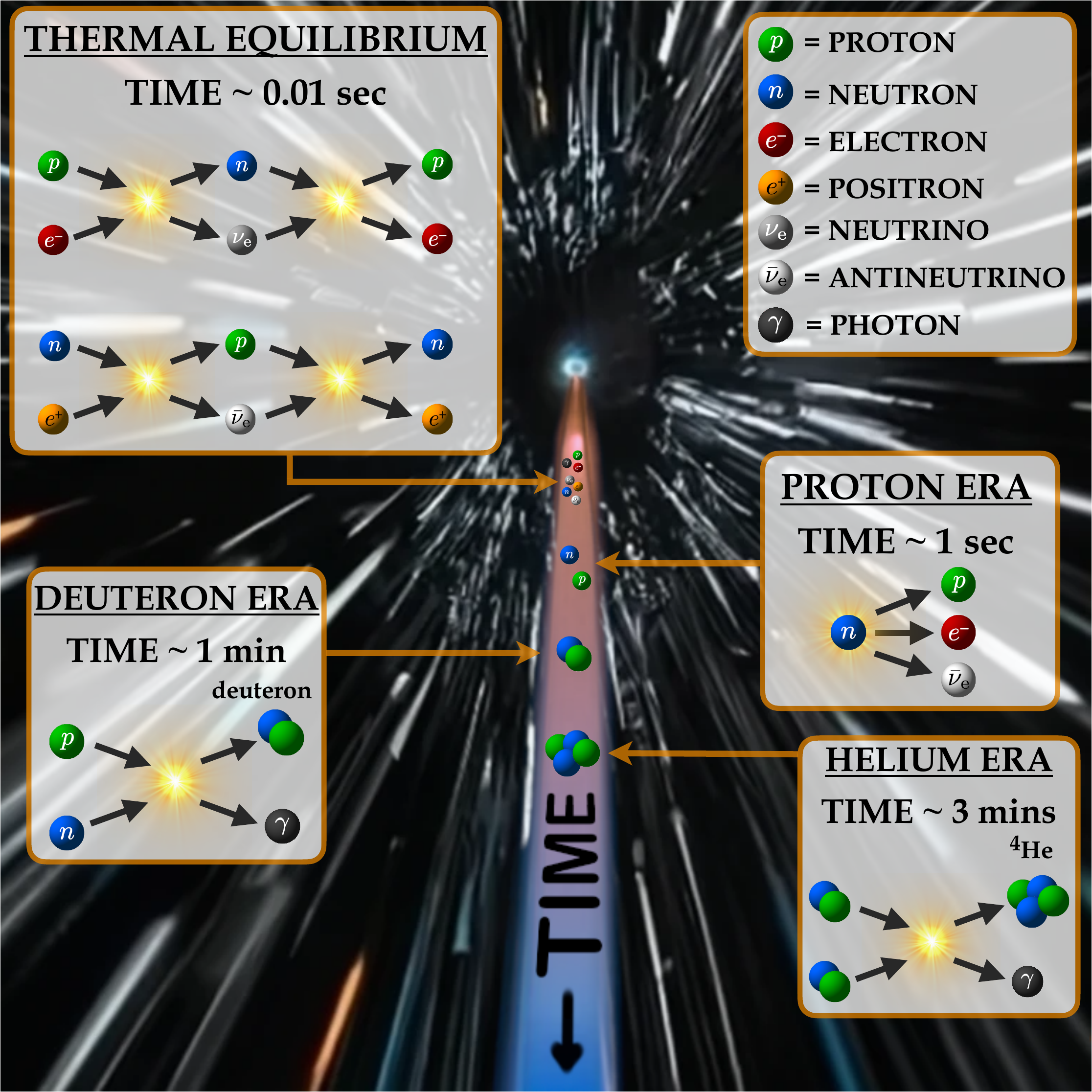}
\end{figure}

One of the most compelling pieces of evidence of the Hot Big Bang model is the realisation and confirmation that some nuclides were created shortly after the Big Bang. This process is referred to as Big Bang nucleosynthesis (or, sometimes, primordial nucleosynthesis), and is the end-product of putting neutrons and protons in a hot, expanding Universe. Big Bang nucleosynthesis currently provides our earliest test of cosmology, and it is the only experiment currently designed that is simultaneously sensitive to all four known fundamental forces: the gravitational force, the electromagnetic force, the strong force and the weak force. Our theoretical understanding of Big Bang nucleosynthesis and the measurement of the primordial abundances together represents one of the strongest pillars of the standard cosmological model. In this chapter, we will develop an intuitive understanding of Big Bang nucleosynthesis, discuss modern calculations of this process, and provide a summary of the current state-of-the-art measurements that have been made. Overall, Big Bang nucleosynthesis is in remarkable agreement with various cosmological probes, and it is this agreement that serves to strengthen our confidence in the general picture of cosmology that we have today.
\end{abstract}

\begin{keywords}
Big Bang nucleosynthesis, cosmological parameters, baryon density, neutrino decoupling, cosmological neutrinos, quasar absorption line spectroscopy, H II regions, stellar abundances.
\end{keywords}

\begin{abstract}[Key points and learning objectives]
\begin{itemize}
    \item Big Bang nucleosynthesis currently provides our earliest probe of the physics of the very early universe.
    \item Learn about the creation of the first nuclides during the first few minutes after the Big Bang.
    \item Develop an understanding of the ingredients that need to be included in a calculation of Big Bang nucleosynthesis.
    \item Obtain a broad overview of the observational techniques currently adopted to infer the abundances of the most common primordial nuclides.
\end{itemize}
\end{abstract}

\section{Introduction: The Ashes of the Big Bang}

\subsection{The development of Big Bang nucleosynthesis}

It was once believed that the first stars formed out of a pure hydrogen gas. Then, nuclear reactions that occurred within the first stars formed some of the first helium nuclei and many of the other heavier elements too. This idea was almost true, however, there was a slight problem with this picture; there were no known processes that were capable of producing the amount of helium that is observed in stars, including the helium abundance of the Sun. This problem was solved around the mid-$20^{\rm th}$ century, with the appreciation that chemical elements could be created during the early Universe when the temperature and density of the primordial plasma were both much higher than they are today \citep{AlpherBetheGamow1948}. The original motivation of this research was to account for the existence of \emph{all} chemical elements, and not just helium. Quite amazingly, this early work correctly estimated the amount of helium that would be produced during an early period of nucleosynthesis, but grossly overestimated the abundances of all other chemical elements because several key physical processes were missing from the calculation (in particular, weak interactions were not included and several unrealistic nuclear cross-sections were assumed in an attempt to reproduce the terrestrial abundances). Nevertheless, this work recognised that there was a brief period of nuclear fusion during the early Universe. The first calculations that included the key processes relevant to the early Universe surfaced some two decades later \citep{HoyleTayler1964,Peebles1966,WagonerFowlerHoyle1967}. These works marked the beginning of what is now known as Big Bang nucleosynthesis (BBN; also referred to as primordial nucleosynthesis).

\begin{figure}
    \centering
    \includegraphics[width=\columnwidth]{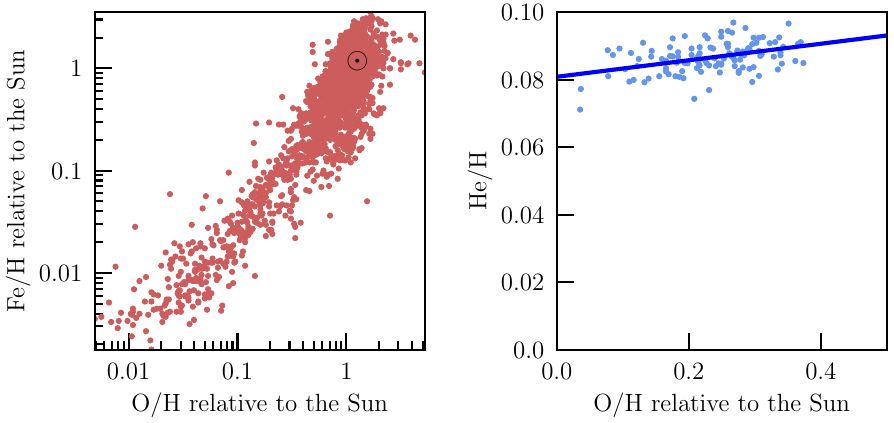}
    \caption{\emph{Left panel:} Each red symbol is a measurement of the number of oxygen atoms (O) relative to hydrogen atoms (H) in stars compared to the number of iron atoms (Fe) relative to hydrogen atoms. Both ratios are relative to the solar value, so a value of one on both axes represents the chemical composition of the Sun (shown as the $\odot$ symbol). Note that the left panel is shown on a logarithmic scale to emphasize that both the O/H and Fe/H ratios closely track each other towards Fe/H=0 when O/H=0.
    \emph{Right panel:} The x-axis is the same as the left panel, but now shown on a linear scale. The y-axis shows the number of helium atoms (He) relative to hydrogen atoms for a sample of star-forming galaxies (blue symbols; see Section~\ref{sec:helium} for more details). The solid blue line is a linear fit to the data, extrapolated to O/H=0. The red data points are from the Stellar Abundances for Galactic Archaeology database \citep{Suda2008}, while the blue data points are from \citep{Hsyu2020}.}
    \label{fig:heliummetals}
\end{figure}

For reasons that we will return to later, BBN is not capable of producing significant quantities of the elements that are heavier than helium. Instead, stars are chiefly responsible for producing most of the chemical elements of the periodic table, and this has been tested empirically. Astronomers have studied a wide range of environments (including stars, galaxies, and gas clouds), and have found that the abundances of almost all of the chemical elements that make up the periodic table decreases as the amount of `metals' in these environments decreases (where the term `metals' generally refers to all chemical elements heavier than helium; see left panel of Figure~\ref{fig:heliummetals} for an example). There are just a few notable exceptions to this, and among the most striking of these is helium (see the right panel of Figure~\ref{fig:heliummetals}), which is the most abundant chemical element in the Universe after hydrogen. If helium was first made by the first stars, then we would expect its abundance to decrease as the abundance of metals decreases, such that at zero metals there would be zero helium. Instead, we find a non-zero abundance of helium when there are no metals. This implies that the first stars were born from a primordial composition that contained a significant amount of both hydrogen and helium.

\subsection{The early Universe --- a modern perspective}
\label{sec:bbnmodern}

Immediately after the Big Bang, the Universe was filled with all known particles (and perhaps some others that we still don't know about!). All of these particles were in thermal equilibrium and moving with almost the speed of light. This brief moment in the history of the Universe involved intense pair creation and annihilation. The expansion of the Universe played a crucial role in bringing an end to this constant battle of creation and annihilation; as the Universe expanded, the radiation temperature dropped, the soup of particles cooled, and interactions became less frequent. Roughly $\sim1~{\rm s}$ after the Big Bang, the only Standard Model particles still remaining were neutrinos, antineutrinos, electrons, positrons, protons, neutrons, and photons. All of these particles were in thermal equilibrium, which means that the interaction rates between particle species were faster than the expansion rate of the Universe. In order to maintain thermal equilibrium, particles need to be `coupled', which means that they constantly interact.

As the Universe continued its expansion, interactions became less frequent, and the neutrinos (and antineutrinos) started to decouple from the other particles. This occurred when the temperature ($T$) of the Universe was roughly $k_{\rm B}\,T\simeq1~{\rm MeV}$ (where $k_{\rm B}$ is the Boltzmann constant), which is about $1~{\rm s}$ after the Big Bang. However, before the neutrinos had a chance to fully decouple from the other particles, electrons and positrons started to annihilate. These final interactions (between neutrinos, electrons and positrons) gently heated the neutrinos just prior to the onset of BBN. This introduced a subtle change to the expansion rate. For now, we will ignore this process, and come back to it in Section~\ref{sec:bbncalc}, once we are armed with a basic understanding of BBN. After most of the positrons found an electron to annihilate with, there was just a small residual number of electrons; since the Universe is observed to be charge-neutral to a high degree of accuracy, there must have existed the same number of protons and electrons at this time. More importantly, after all particle annihilation had taken place, there was a slight residual abundance of baryons over anti-baryons; this residual of baryons is referred to as the baryon asymmetry, the origin of which is not fully understood. As we will see in the following sections, baryons (specifically neutrons and protons) are the main actors in BBN, and it is the abundance of the baryons that largely determines the creation of the first nuclides. After the majority of the electrons and positrons had annihilated, but just prior to BBN, the Universe contained decoupled neutrinos, electrons, protons, neutrons and photons (and, for completeness, there is also a negligible number of positrons remaining that will eventually find an electron within about 2 hours after the Big Bang, and subsequently annihilate; \citealt{Thomas2020}).

\subsection{Genesis of the first elements}
\label{sec:bbntheory}

With the above knowledge, we are now in a position to consider a simplified analytical calculation of BBN. The first step is to determine the initial conditions of the problem, and this requires us to calculate the relative abundance of neutrons and protons. We start by considering the state of the Universe prior to neutrino decoupling. As discussed earlier, regular interactions between particles maintain thermal equilibrium at this time  (recall, this means that the interaction rates are much higher than the expansion rate of the Universe). In thermal equilibrium, the number density of a particle is given by:
\begin{equation}
    N=g\,\bigg(\frac{m\,k_{\rm B}\,T}{2\pi\,\hbar^{2}}\bigg)^{3/2}\,\exp\bigg[\frac{\mu-m\,c^{2}}{k_{\rm B}\,T}\bigg]
\end{equation}
where $g$ is the number of internal degrees of freedom, $m$ and $\mu$ are the mass and chemical potential of the particle, $\hbar$ is the reduced Planck constant, and $c$ is the speed of light. The relative numbers of neutrons and protons are determined by the following charged-current weak interactions:
\begin{eqnarray}
    \label{eqn:ccwia}
    n+e^{+} &\longleftrightarrow& p+\Bar{\nu}_{e}\\
    p+e^{-} &\longleftrightarrow& n+\nu_{e}\\
    \label{eqn:ccwib}
    n &\longleftrightarrow& p+e^{-}+\Bar{\nu}_{e}
\end{eqnarray}
where the symbols take on the following meanings: $n={\rm neutron}$, $p={\rm proton}$, $e^{+}={\rm positron}$, $e^{-}={\rm electron}$, $\Bar{\nu}_{e}={\rm electron~antineutrino}$, $\nu_{e}={\rm electron~neutrino}$. If these rates are rapid compared to the expansion of the Universe, then the primordial plasma is also in chemical equilibrium, and we have $\mu_{n}+\mu_{\nu_{e}}=\mu_{p}+\mu_{e^{-}}$. In this situation, we can write down the number density of neutrons relative to protons:
\begin{equation}
    \label{eqn:np}
    N_{\rm n}/N_{\rm p} = (m_{\rm n}/m_{\rm p})^{3/2}\,\exp\bigg[-\frac{(m_{\rm n}-m_{\rm p})\,c^{2}}{k_{\rm B}\,T}\bigg]
\end{equation}
where the difference of the chemical potentials $\mu_{n}-\mu_{p}\equiv\mu_{e^{-}}-\mu_{\nu_{e}}$ does not appear in the above equation, since it is close to zero. This is because (to first order) $\mu_{e^{-}}/{k_{\rm B}\,T} \sim N_{e^{-}}/N_{\gamma}$, where $N_{\gamma}$ is the photon number density. Given that the Universe is charge neutral, we expect this ratio to be close to the ratio of baryons to photons (i.e. $\sim N_{p}/N_{\gamma}\lesssim10^{-9}$). A similar argument holds for $\mu_{\nu_{e}}$.

For the purpose of this exercise, the term $(m_{\rm n}/m_{\rm p})\sim1$, while the term in the exponential $(m_{\rm n}-m_{\rm p})\,c^{2}=1.293\,{\rm MeV}$. Therefore when the temperature of the Universe ${k_{\rm B}\,T}\gg1\,{\rm MeV}$, there are roughly the same number of neutrons and protons. As the Universe expands, the temperature drops, so that there are more protons than neutrons when ${k_{\rm B}\,T}<1\,{\rm MeV}$. Thus, if the weak interaction rates (Equations~\ref{eqn:ccwia}--\ref{eqn:ccwib}) were efficient, then the neutron/proton ratio would continue to drop as the Universe expands and cools. However, the reaction rate per neutron is of the form $\Gamma=n_{\nu_{\rm e}}\,c\,\langle\sigma\rangle$, where $n_{\nu_{\rm e}}$ is the number density of electron neutrinos (which is proportional to $T^{3}$), and the weak-interaction cross-section in this temperature regime $\langle\sigma\rangle\propto T/\tau_{n}$, where $\tau_{n}\sim880\,{\rm s}$ is the mean lifetime of a neutron (note, the half-life of a neutron is $\sim610\,{\rm s}$). Thus, the interaction rate falls off as $T^{4}$ as the Universe expands. The expansion rate during the early Universe is dominated by radiation, where the radiation density has a temperature dependence $\rho\propto T^{4}$. The Friedmann equations tells us that the expansion rate is given by:
\begin{equation}
    \frac{\dot{a}}{a}\equiv H = \sqrt{\frac{8\pi\,G\,\rho}{3}}\propto T^{2}
\end{equation}
where $a$ is the scale factor of the Universe, $H$ is the Hubble parameter, $G$ is the Gravitational constant, and $\rho$ is the volume mass density. Thus, the weak interaction rates ($\propto T^{4}$) suddenly drop below the expansion rate of the Universe ($\propto T^{2}$) as the Universe cools; the weak interaction rates cease to be important around ${k_{\rm B}\,T}\simeq0.8\,{\rm MeV}$. Inserting this value into Equation~\ref{eqn:np}, gives a `freeze-out' neutron abundance of $N_{\rm n}/N_{\rm p}\sim1/5$. This is the initial condition of BBN.

At this point, the Universe is mostly just protons and neutrons locked in a ratio of $5:1$. Because of the proton coulomb barrier, $p+p$ nuclear reactions are not favourable. Furthermore, $n+n$ reactions are also not favourable, because the spins of the neutrons need to be the opposite of each other (due to the Pauli exclusion principle) and the nuclear binding is much weaker when the neutron spins are different. Instead, the first nuclear reaction rate involves the synthesis of the deuteron, which is the bound state of a proton and neutron. The deuteron is the nucleus of a deuterium atom, which is an isotope of the hydrogen atom. The reaction rate to produce (and destroy) deuterons is:
\begin{equation}
    \label{eqn:npd}
    n+p \longleftrightarrow d+\gamma
\end{equation}
where $d$ represents the deuteron and $\gamma$ is a photon. Note that this reaction proceeds in both directions; deuterons are created by this process, but can also be destroyed by photons and thereby produce one neutron and one proton. In order for there to be net deuteron production, we need the creation rate to be much larger than the destruction rate. The binding energy of a deuteron is ${k_{\rm B}\,T}=2.224\,{\rm MeV}$, so deuterons do not form in abundance until the temperature of the photons drop well below the deuteron binding energy. In fact, because baryons are so underabundant --- by a factor of $\sim10^{9}$ --- compared to the photons, deuterons are destroyed by the high energy tail of the blackbody distribution until the photon temperature drops to well below the deuteron binding energy. This is known as the `deuteron bottleneck' (and sometimes incorrectly as the `deuterium bottleneck'). The creation and destruction rates of Equation~\ref{eqn:npd} are equal when the Universe is at a temperature of ${k_{\rm B}\,T}\simeq0.06\,{\rm MeV}$, and at this point deuteron production begins in earnest.

Now that the Universe has grown in complexity (i.e. BBN can now proceed with neutrons, protons, \textit{and} deuterons!), several other nuclear reaction rates become important. The most important reactions during the earliest phases of BBN include those that synthesize \het\ (a stable nuclide containing two protons and one neutron):
\begin{eqnarray}
    d + p &\longrightarrow& ^{3}{\rm He} + \gamma\\
    d + d &\longrightarrow& ^{3}{\rm He} + n \label{eqn:ddn}
\end{eqnarray}
as well as reactions that can make triton, $^{3}$H (an unstable nuclide containing one proton and two neutrons --- an isotope of hydrogen):
\begin{eqnarray}
    d + d &\longrightarrow& ^{3}{\rm H} + p  \label{eqn:ddp} \\
    ^{3}{\rm He} + n &\longrightarrow& ^{3}{\rm H} + p
\end{eqnarray}
and finally, several reactions that are capable of synthesising \hef\ (a stable nuclide containing two protons and two neutrons; sometimes referred to as $\alpha$ particles):
\begin{eqnarray}
    \label{eqn:hefa}
    d + d &\longrightarrow& ^{4}{\rm He} + \gamma\\
    d + {^{3}{\rm He}} &\longrightarrow& ^{4}{\rm He} + p\\
    \label{eqn:hefc}
    d + {^{3}{\rm H}} &\longrightarrow& ^{4}{\rm He} + n
\end{eqnarray}
The above reactions are the most important reaction rates during the initial stages of BBN. Once a considerable abundance of \hef\ is produced, several mass-7 nuclides can also be formed via the following channels:
\begin{eqnarray}
    ^{3}{\rm H} + {^{4}{\rm He}} &\longrightarrow& ^{7}{\rm Li} + \gamma\\
    ^{3}{\rm He} + {^{4}{\rm He}} &\longrightarrow& ^{7}{\rm Be} + \gamma\\
    ^{7}{\rm Be} + n &\longrightarrow& ^{7}{\rm Li} + p\\
    ^{7}{\rm Li} + p &\longrightarrow& 2~{^{4}{\rm He}}
\end{eqnarray}
Of course, we could continue by writing down a huge network of nuclear reactions that can affect the outcome of BBN, but this list includes the most important nuclear reaction rates that determine the order of magnitude abundances of the primordial nuclides. An accurate determination of the abundances requires a detailed numerical calculation (see Section~\ref{sec:bbncalc}). For the moment, we can continue with a simplified calculation of BBN.

If the above nuclear reaction rates are much faster than the expansion rate of the Universe, then the nuclides will be in nuclear statistical equilibrium (NSE), and their abundances will be proportional to $\propto\exp[E/k_{\rm B}\,T]$, where $E$ is the binding energy of the nucleus. The most tightly bound light element that is involved in the above reactions is \hef, with a binding energy $E\sim28.3\,{\rm MeV}$. The next most important nuclides are \het\ and $^{3}$H, which have binding energies a factor of $\sim4$ lower than \hef; this factor of $\sim4$ is enormous in the exponential. Since BBN is very close to NSE, the dominant nuclide synthesized is therefore \hef. We can estimate the number abundance of \hef\ by assuming that all neutrons that are available for nucleosynthesis are incorporated into \hef.

To begin this calculation, we can use the initial abundance of neutrons relative to protons (1:5) that we calculated earlier in this chapter. Recall that this ratio is set prior to the onset of BBN, when ${k_{\rm B}\,T}\simeq0.8\,{\rm MeV}$. Furthermore, \hef\ can only be made by two-body reactions (see Equations~\ref{eqn:hefa}--\ref{eqn:hefc}), and these reactions require either $d$, \het, or $^{3}{\rm H}$ (the latter two require $d$) as fuel to make \hef. As a result, \hef\ can only attain the NSE value when sufficient deuterons are available. As we found earlier, this occurs when the Universe is at a temperature of ${k_{\rm B}\,T}\simeq0.06\,{\rm MeV}$, roughly $\sim5$ minutes ($=300\,{\rm s}$) after the neutrons `freeze-out' at ${k_{\rm B}\,T}\simeq0.8\,{\rm MeV}$. During this time, free neutrons will beta decay (see Equation~\ref{eqn:ccwib}), reducing the number of neutrons available for the nucleosynthesis of \hef. The fraction of neutrons that have been converted into protons is given by $\exp[-t/\tau_{n}]\simeq\exp[-300/880]\simeq0.7$. Thus, the number of neutrons decreases by $\sim30\%$ and this slightly increases the number of protons. The final number density of neutrons is therefore $n_{\rm f}=0.7\,n_{\rm i}$, while the final number density of protons is $p_{\rm f}=p_{\rm i} + 0.3\,n_{\rm i}$, where the subscripts $i$ refer to the initial number densities. Therefore, the ratio of neutrons to protons around the time that \hef\ is synthesized is $n_{\rm f}/p_{\rm f}\sim0.14$. Since all neutrons end up in \hef, which contains two neutrons, the number density of \hef\ is $N(^{4}{\rm He})=n/2$ and the number density of hydrogen nuclides is $N(^{1}{\rm H})=p-n$, leading to a primordial helium number abundance:
\begin{equation}
    \frac{N(^{4}{\rm He})}{N(^{}{\rm H})} \equiv y_{\rm P} \simeq 0.08
\end{equation}
where the variable $y_{\rm P}$ is commonly used to represent the primordial number density of \hef\ relative to H. This relatively simple BBN calculation is remarkably close to the value that is inferred from observational studies (e.g. the right panel of Figure~\ref{fig:heliummetals}). It is also common to see the variable $Y_{\rm P}$ (note the uppercase `Y') used in the literature. This form was used historically, and represents the \emph{mass fraction} of \hef. The mass fraction is related to the number abundance by the following equation:
\begin{equation}
    Y_{\rm P}\coloneqq\frac{4\,y_{\rm P}}{1+4\,y_{\rm P}}
\end{equation}
For $y_{\rm P} \simeq 0.08$, the corresponding mass fraction is $Y_{\rm P} \simeq 0.24$, which implies that \hef\ makes up roughly one-quarter of all baryons by mass.

The assumption that all neutrons are incorporated in \hef\ is quite accurate, but not entirely true. Some neutrons are incorporated into the lighter nuclides (d, \het, $^{3}$H), while some neutrons end up in heavier nuclides ($^{6}$Li, $^{7}$Li, $^{7}$Be, among many more!). Indeed, some neutrons never find a nucleus, and instead beta decay to raise the number of protons, electrons and electron antineutrinos by an insignificant amount (see Equation~\ref{eqn:ccwib}).

An important aspect of BBN is that there are no stable nuclides with a mass number of 5 or 8. This restricts the formation of primordial nuclides to the lowest atomic number elements, as heavier elements are not capable of being made in appreciable abundance. For example, $^{12}$C has a binding energy $E\sim92.2\,{\rm MeV}$, more than three times that of \hef. If it were possible to form $^{12}$C during BBN, it would have been preferred over \hef. However, the primary formation channel of $^{12}$C is the triple-alpha process: First, $^{8}$Be is made via the fusion of two \hef\ nuclides, then a third \hef\ nuclide needs to fuse with $^{8}$Be in less than $8.2\times10^{-17}\,{\rm s}$, which is the half-life of an unstable $^{8}$Be nucleus. The relatively low particle density during BBN, combined with the relatively short period of time that BBN occurs, means that this reaction is not able to produce a significant quantity of $^{12}$C; in other words, the expansion rate is much faster than the triple-alpha process. The fact that there are no stable nuclides with a mass number of 5 or 8, combined with \hef\ having a much larger binding energy than any of the other nuclides with mass number $\le7$, are the reasons why \hef\ is the dominant nuclide created during BBN.

\section{Calculating the creation of the first nuclides}
\label{sec:bbncalc}

The aforementioned `simplistic' approach to studying BBN can only provide us with a grasp of the key concepts, and provide some intuition about the processes that are taking place during the first few minutes after the Big Bang. In order to use BBN for precision cosmology, we need to perform a numerical calculation that solves the coupled Boltzmann equations, includes a full nuclear reaction network, and all of the relevant physical processes. Before charging onwards, let us first recap some of the most important points outlined in Section~\ref{sec:bbntheory}: BBN is fundamentally a competition between the nuclear reaction rates, the free neutron half-life, and the expansion rate of the Universe. The reaction rates depend on the density of the reactants (i.e. the nuclides) and their energy (i.e. the temperature of the plasma). Since the density of the reactants also changes with the expansion rate of the Universe, it is convenient to normalize the nucleon density by the photon density. This important quantity is referred to as the `baryon-to-photon ratio', and is usually denoted $\eta=n_{\rm B}/n_{\rm \gamma}$, where $n_{\rm B}$ and $n_{\rm \gamma}$ are the baryon and photon number densities, respectively. It is the ratio of baryons to photons that determines the creation and destruction of the primordial nuclides, and is (very nearly) a conserved quantity as the Universe expands.

The expansion rate of the Universe is determined by the content of the Universe (i.e. through the Friedmann equations). During the earliest periods of the Universe, the total energy density is dominated by relativistic particles (i.e. `radiation', $\rho_{R}$), so the expansion rate is of the form:
\begin{equation}
    H(T)^{2} = \frac{8\,\pi\,G}{3}\rho_{R}(T)
\end{equation}
where $\rho_{R}(T)$ is the total radiation density as a function of the plasma temperature. BBN codes usually track the time ($\equiv$ thermal) evolution of the expansion rate from long before electron/positron annihilation ($E=k_{\rm B}T\gtrsim$ a few MeV), until long after the freeze-out of the BBN abundances ($E\ll m_{\rm e}$, where $m_{\rm e}$ is the rest mass of an electron). Just prior to neutrino decoupling, the only Standard Model particles that are present include:
\begin{itemize}
    \item photons ($g_{\gamma}=2$). Recall that $g$ is the number of internal degrees of freedom, where photons have two polarisations, so $g=2$.
    \item electron/positron pairs ($g_{\rm e}=4$). Both electrons and positrons can have spin up or down, so $g=2\times2$.
    \item neutrinos/antineutrinos ($g_{\nu}=2N_{\nu}=6$). The Standard Model contains three neutrino species and their corresponding antiparticles. Neutrinos are always left-handed ($g=3$) and antineutrinos are always right-handed ($g=3$).
\end{itemize}
BBN calculations therefore start with the assumption, $\rho_{R}=\rho_{\gamma} + \rho_{e} + \rho_{\nu}$ where all particles are in thermal equilibrium. Expressed relative to the photon density, we have:
\begin{equation}
\label{eqn:rhorgamma}
    \frac{\rho_{R}}{\rho_{\gamma}} = 1 + \frac{\rho_{e}}{\rho_{\gamma}} + \frac{\rho_{\nu}}{\rho_{\gamma}} = 1 + \frac{7}{8}\bigg[ \frac{g_{\rm e}}{g_{\gamma}} + \frac{g_{\nu}}{g_{\gamma}} \bigg] = \frac{43}{8}
\end{equation}
where the factor $7/8$ comes from the ratio of the energy density of particles obeying Fermi-Dirac statistics (applicable to electrons, positrons and neutrinos) to particles that obey Bose-Einstein statistics (applicable to photons). Deviations from the Standard Model expansion rate are usually parameterized by an `effective number of neutrino species', $N_{\rm eff}=N_{\nu}+\Delta N_{\nu}$. It is noteworthy that $\Delta N_{\nu}$ need not have anything to do with neutrinos; it is simply a means to parameterize deviations from the Standard Model expansion rate. Thus, in a Universe with a `non-standard' radiation energy density ($\rho_{R}'$), we can rewrite Equation~\ref{eqn:rhorgamma} as:
\begin{equation}
\label{eqn:rhorgammab}
    \frac{\rho_{R}'}{\rho_{\gamma}} = \frac{43}{8} + \frac{7}{8}\Delta N_{\nu}
\end{equation}
Soon after, the neutrinos decouple from the other particles, but still maintain a thermal distribution with a temperature identical to that of the photons, since the expansion of the Universe decreases both the photon and neutrino temperature by an equivalent amount. Then, the electrons and positrons annihilate and transfer their heat and entropy to the photons, thereby raising the photon temperature. If we assume that entropy ($s\propto g\,T^{3}$) is conserved before and after electron/positron annihilation, and that electron/positron annihilation is the only source of heating, then we can estimate the temperature of the cosmic neutrino background relative to the cosmic photon background (a.k.a. the cosmic microwave background, [CMB]):
\begin{equation}
    \frac{T_{\nu}}{T_{\gamma}}=\bigg(\frac{g_{\gamma}}{g_{\gamma}+(7/8)g_{\rm e}}\bigg)^{1/3} = \bigg(\frac{4}{11}\bigg)^{1/3}
\end{equation}
Thus, the temperature of the cosmic neutrino background is $\sim30\%$ lower than the CMB temperature. Once most of the electrons and positrons have annihilated, the radiation density then becomes:
\begin{equation}
    \frac{\rho_{R}}{\rho_{\gamma}} = 1 + \frac{\rho_{\nu}}{\rho_{\gamma}} =
    1 + \frac{7}{8}N_{\nu}\bigg(\frac{T_{\nu}}{T_{\gamma}} \bigg)^{4} =
    1 + \frac{7}{8}N_{\nu}\bigg(\frac{4}{11} \bigg)^{4/3}
\end{equation}
or, in a Universe with a `non-standard' expansion rate parameterized by $N_{\rm eff}$, we have:
\begin{equation}
\label{eqn:rhodgamma}
    \frac{\rho_{R}'}{\rho_{\gamma}} = 1 + \frac{7}{8}\bigg(\frac{4}{11} \bigg)^{4/3}\big(3 + \Delta N_{\nu}\big)
\end{equation}
Note that Equation~\ref{eqn:rhodgamma} assumes that neutrino decoupling is well-separated from electron/positron annihilation. In reality, electron/positron pairs start annihilating before the highest energy neutrinos are fully decoupled, and this overlap gently heats the neutrinos. Detailed calculations of this process have shown that $N_{\rm eff}=3.043$ for the Standard Model ($N_{\nu}=3$; \citealt{Cielo2023}). Thus, the term in parentheses in Equation~\ref{eqn:rhodgamma} should be replaced with ($3.043+\Delta N_{\nu}$). Similarly, the term $(43/8)$ in Equations~\ref{eqn:rhorgamma} and \ref{eqn:rhorgammab} should be replaced with $(22/8 + 7\times3.043/8)$. To be clear, after applying this small correction, $N_{\rm eff}=3.043$ corresponds to $\Delta N_{\nu}=0$. Any deviation from $N_{\rm eff}\simeq3.043$ (i.e. $\Delta N_{\nu}\neq0$) would indicate physics that is not currently captured by our modern perspective of BBN.

\begin{figure}
    \centering
    \includegraphics{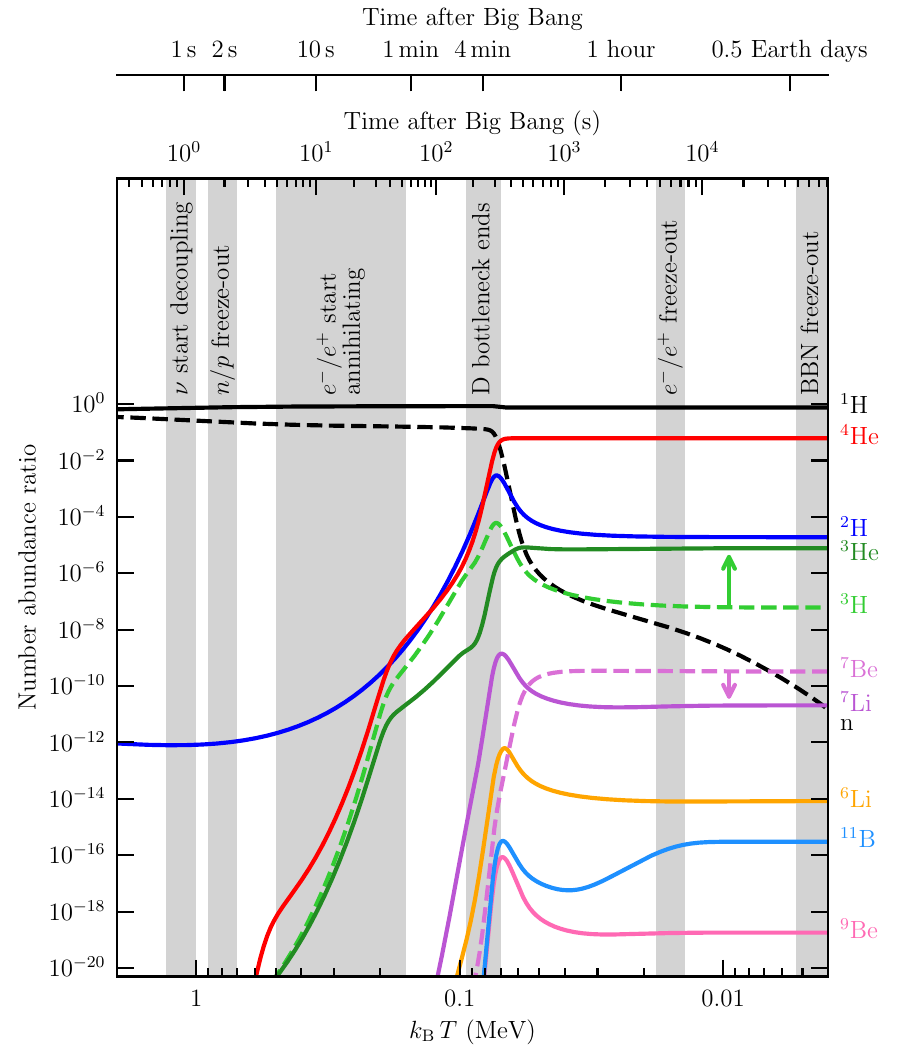}
    \caption{The time evolution of the primordial nuclides, as labelled on the right side of the figure. The bottom x-axis shows the temperature evolution of the Universe, while the two top x-axes indicate the corresponding time that has elapsed since the Big Bang. The neutrons and protons are in equilibrium at the earliest times, until the weak interaction rates fall below the expansion rate of the Universe ($\sim2\,{\rm s}$ after the Big Bang), and the relative abundance of neutrons and protons freezes out. Free neutrons (dashed black curve) start decaying, and continue to decay throughout and beyond the period of BBN, until there are no free neutrons remaining. The lowest energy neutrinos start to decouple from the other particles at around $\sim1\,{\rm s}$ after the Big Bang, and this process continues until some of the highest energy neutrinos decouple during the beginning of $e^{-}/e^{+}$ annihilation. $e^{-}$ and $e^{+}$ continue to annihilate throughout BBN, until the expansion rate exceeds the $e^{-}$ annihilation rate when the temperature of the Universe falls below $\sim 16~{\rm keV}$ ($\sim 2~{\rm hours}$ after the Big Bang). Meanwhile, the lightest nuclides, particularly deuterons (dark blue curve), start forming in appreciable abundance when the photon temperature drops below $\sim 0.1~{\rm MeV}$, at which point \hef\ can establish its nuclear statistical equilibrium abundance (red curve). Most of the primordial nuclides freeze-out around $8$\,hours after the Big Bang; when the Universe is $\lesssim6~{\rm keV}$, the primordial nuclides change from their final abundances by $<0.1\%$. However, some unstable nuclides (e.g. $^{3}$H and $^{7}$Be, shown as dashed lines) will decay long after the BBN abundances freeze out. Arrows are shown from each unstable BBN nuclide to its eventual decay product.}
    \label{fig:abundance_time}
\end{figure}

Armed with an understanding of the expansion rate, the nuclear reaction rates, and the mean lifetime of a free neutron, we are now in a position to perform a complete calculation of BBN. The only free parameter is the baryon-to-photon ratio ($\eta$), which is related to the present day baryon density of the Universe ($\Omega_{\rm B,0}\equiv\rho_{\rm B}/\rho_{\rm crit}$) by the equation \citep{Steigman2006}:
\begin{equation}
    \eta = (273.78\pm0.18)~\times~10^{-10}~\times~\Omega_{\rm B,0}\,h^{2}
\end{equation}
where $h\coloneqq H_{0}/100~{\rm km~s}^{-1}~{\rm Mpc}^{-1}$ and $H_{0}$ is the present day Hubble constant. If we adopt the $\Omega_{\rm B,0}\,h^{2}$ value derived from the CMB, BBN becomes a parameter free theory. The results of an example calculation \citep[based on the code described by][]{Gariazzo2022}, are shown in Figure~\ref{fig:abundance_time}, where the abundances of several primordial nuclides are monitored during the first 24 hours after the Big Bang, and several key moments around the time of BBN are labelled. Recall that only the lightest elements of the periodic table were made appreciably during BBN, since there are no stable nuclides with a mass number of 5 or 8. The first nuclides to freeze-out (i.e. change by $\lesssim0.1\%$ relative to the final state of the Universe) are $^{1}$H and $^{4}$He, which are also the most abundant of the light nuclides. This freeze-out occurred $\sim6$ minutes after the Big Bang. Of the nuclides shown in Figure~\ref{fig:abundance_time}, $^{7}$Be freezes-out next, at around $\sim4$ hours after the Big Bang, while $^{2}$H, $^{3}$H, $^{3}$He, and $^{7}$Li all freeze-out at roughly the same time, around $\sim8-10$ hours after the Big Bang.

Technically, the above `freeze-out' abundances refer to the freeze-out of nuclear reactions. The quoted freeze-out times do not include the decay of the unstable primordial nuclides. As mentioned earlier, free neutrons will decay via Equation~\ref{eqn:ccwib}, with a mean lifetime of $\tau_{\rm n}\sim880~{\rm s}$. Triton is next to decay, with a mean lifetime of $\tau\sim12.32$~years, via $^{3}{\rm H} \to {^{3}{\rm He}} + e^{-} + \bar{\nu}_{e}$. Finally, of the remaining nuclides shown in Figure~\ref{fig:abundance_time}, $^{7}$Be begins decaying to $^{7}$Li via electron-capture ($^{7}{\rm Be} + e^{-} \to {^{7}{\rm Li}} + \nu_{e}$) starting at a redshift $z\sim35,000$ (i.e. $\sim600$ years after the Big Bang; \citealt{Khatri2011}), and this process continues until $z\sim27,500$ (i.e. $\sim1000$ years after the Big Bang). For all intents and purposes, it is extremely unlikely that the abundances of primordial $^{3}{\rm H}$ and $^{7}{\rm Be}$ will ever be measured directly. Instead, it is common to assume that the primordial production of lithium-7 is the sum of the primordially produced ${^{7}{\rm Li}}+{^{7}{\rm Be}}$. Similarly, the final amount of helium-3 that is produced is quoted as the sum of the primordially produced ${^{3}{\rm H}}+{^{3}{\rm He}}$.

\begin{figure}
    \centering
    \includegraphics[width=\columnwidth]{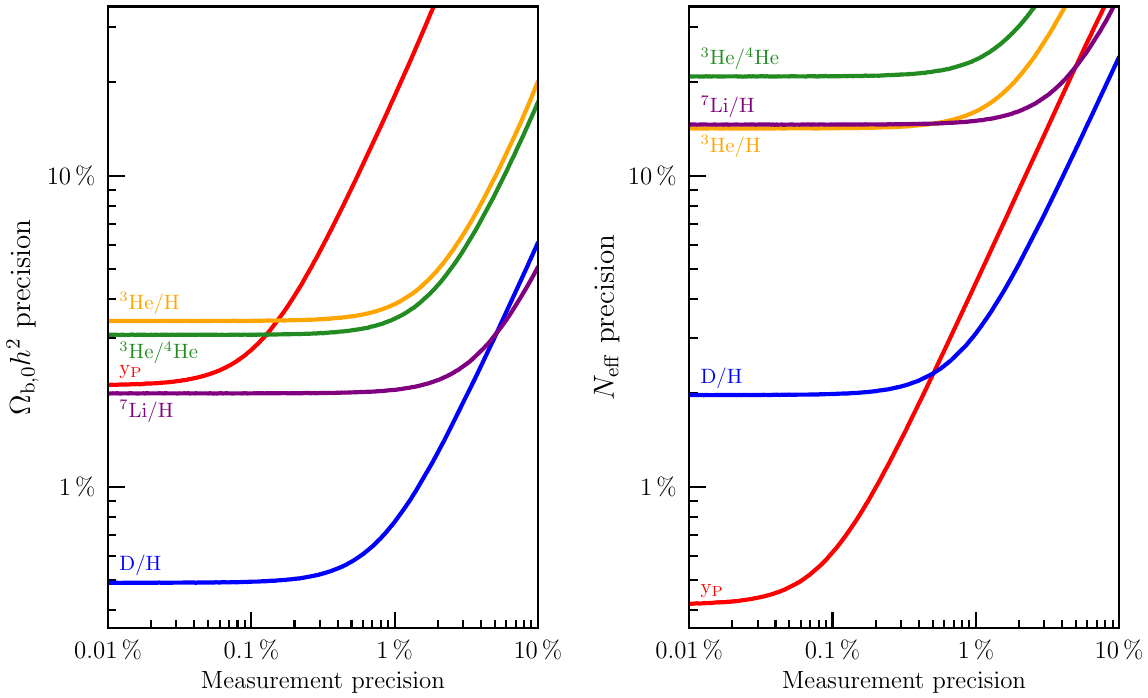}
    \caption{The sensitivity of several primordial abundance ratios to the baryon density (left panel) and effective number of neutrino species (right panel). The Standard Model is assumed in both panels, while allowing either $\Omega_{\rm B,0}\,h^{2}$, or $N_{\rm eff}$ (but not both simultaneously) to be a free parameter. Each curve reaches a plateau towards the left side of each panel, indicating the limiting precision that is possible given the uncertainties associated with BBN calculations. On the right side of both panels, each curve approaches a diagonal asymptote, which indicates the intrinsic sensitivity of a given nuclide to either the baryon density or the effective number of neutrino species. Of the nuclide ratios shown, D/H has strong sensitivity to both $\Omega_{\rm B,0}\,h^{2}$ and $N_{\rm eff}$. However, $y_{\rm P}$ can deliver higher precision on $N_{\rm eff}$, given that the input physics used for the BBN calculations (particularly the neutron mean lifetime) is relatively well-determined compared with the input physics that affects the production of the other primordial nuclides.}
    \label{fig:bbn_theory_obs}
\end{figure}

Until now, we have only considered the Standard Model of particle physics and cosmology, and adopted a baryon-to-photon ratio derived from observations of the CMB. We can also consider potential departures from the Standard Model, such as a baryon-to-photon ratio that is different between the times of BBN and the CMB, or a non-standard expansion rate parameterized by $\Delta N_{\nu}\neq0$. An exhaustive investigation of various extensions to the Standard Model is not the goal of this Chapter, but we can still obtain an intuitive understanding of the sensitivity of each primordial abundance to new physics. Consider the case where the expansion rate is identical to the Standard Model, and the goal is to use BBN to determine the baryon density independently of the CMB. Accounting for the uncertainties associated with the calculation (i.e. the nuclear reaction rates, the mean neutron lifetime, etc.), the left panel of Figure~\ref{fig:bbn_theory_obs} shows the 
precision with which $\Omega_{\rm B,0}\,h^{2}$ can be determined from the most abundant
primordial nuclides, given a range of measurement precision. The left asymptote on this plot represents the limiting precision due to the uncertainty associated with BBN calculations; the left asymptote tells us the best precision that is possible with current calculations. The right asymptote represents the intrinsic sensitivity of each primordial abundance to $\Omega_{\rm B,0}\,h^{2}$. Thus, $^{7}$Li/H is the most sensitive primordial abundance to the baryon density, closely followed by D/H. However, the nuclear reactions that are most relevant to deuteron nucleosynthesis are known to better precision than that of $^{7}$Li. The D/H ratio is therefore more suitable to determine the baryon density at high precision than any other primordial nuclide.

Now consider the possibility that the baryon density does not change from BBN to the CMB, but the expansion rate during BBN deviates from the Standard Model expansion rate. The resulting sensitivities of the primordial nuclides to deviations of the expansion rate (assuming the baryon density is known perfectly) are shown in the right panel of Figure~\ref{fig:bbn_theory_obs}. In this case, the D/H ratio is the most sensitive primordial nuclide to the expansion rate of the Universe (parameterized by $N_{\rm eff}$). However, the uncertainty of nuclear reaction rates (particularly the reactions written in Equations~\ref{eqn:ddn} and \ref{eqn:ddp}) currently limit the precision of $N_{\rm eff}$ to $\sim2\%$, based on D/H alone. To obtain sub-percent precision on the effective number of neutrino species during BBN, the optimal choice is the primordial helium abundance, which in principle can deliver a precision of $\sim0.5\%$, equivalent to $\Delta N_{\nu}\simeq0.015$. It is also noteworthy that the primordial helium abundance is relatively insensitive to the baryon density, making it an important probe of deviations from the Standard Model expansion rate.

Overall, Figure~\ref{fig:bbn_theory_obs} demonstrates that each primordial nuclide has a different sensitivity to the baryon density and the effective number of neutrino species. More generally, any model that considers an extension to the Standard Model will cause specific changes to the primordial abundances --- a unique fingerprint. By measuring multiple primordial abundances, there is an opportunity to not only find evidence of new physics, but also to discern what that new physics might be. It is therefore the joint analysis of multiple primordial abundance measurements, combined with detailed calculations and laboratory measured nuclear reaction rates, that provides the strongest test of the Standard Model during the first moments after the Big Bang.

\section{Observing the first nuclides}

Numerical calculations of the primordial nuclides have allowed us to understand the sensitivity of each primordial nuclide to the physics of the early Universe. For example, if there were an additional particle outside of the Standard Model, this might change the expansion rate of the Universe or interact with baryons in a way that changes the production and destruction of the primordial nuclides. Turning this statement around, if we could measure the relative abundances of the primordial nuclides, and combine these measures with a calculation of BBN, then we can learn about particle physics and cosmology a few minutes after the Big Bang. The main challenge is to find environments in the Universe that have not significantly altered the primordial nuclides from the initial abundances that were set a few minutes after the Big Bang. This usually means that we need to find environments that have very few of the elements made by stars (e.g. oxygen, iron).

The usual goal of observational BBN is to measure the number density of one primordial nuclide relative to the number density of protons (i.e. hydrogen nuclei) that are left over after BBN. However, at least in principle, the ratio of any two primordial nuclides is sensitive to the underlying physics of BBN. Even though a huge range of nuclides are made during BBN, only the most abundantly produced primordial nuclides can currently be measured, including: deuterium (D), helium-3 (\het), helium-4 (\hef), and lithium-7 (\li). Let us now discuss the current techniques and measurements of each of these primordial nuclides in turn.

\subsection{Deuterium --- A way to weigh the Universe}
\label{sec:deuterium}

Deuterium is a heavy stable isotope of the hydrogen atom. The additional neutron in the nucleus of a deuterium atom causes the electron energy levels to be shifted relative to that of hydrogen. This is both a blessing and a curse; if we want to measure the relative number of deuterium and hydrogen atoms in a pristine environment, we need to detect the atomic transitions of both deuterium and hydrogen. The best way to detect different atomic transitions is to use a spectrograph that measures the flux density of photons as a function of wavelength. The spectrograph will produce a spectrum, and we can use this spectrum to identify the transitions corresponding to the hydrogen and deuterium atomic transitions. Let's start by estimating the difference between the deuterium and hydrogen atomic energy levels, by considering the Bohr model of the atom. The energy levels based on this model are given by:
\begin{equation}
    \frac{1}{\lambda} = R\,\bigg(\frac{1}{n_{f}^2} - \frac{1}{n_{i}^{2}}\bigg)
\end{equation}
where $\lambda$ is the wavelength of the transition between levels $n_{f}$ and $n_{i}$, and $R$ is the Rydberg constant of that atom. Since deuterium has an extra neutron in the nucleus, this changes the centre of mass of the atom, so the Rydberg constant is slightly different for deuterium and hydrogen. To first order, the ratio of the deuterium and hydrogen wavelengths is therefore just the ratio of their Rydberg constants, which is simply the ratio of the reduced masses:
\begin{equation}
    \frac{\lambda_{D}}{\lambda_{H}} = \frac{R_{\rm H}}{R_{D}} = \bigg(1 + \frac{m_{e}}{m_{d}}\bigg)~\bigg/~\bigg(1 + \frac{m_{e}}{m_{p}}\bigg)
\end{equation}
where $m_{e}$, $m_{p}$, and $m_{d}$ are the rest masses of the electron, proton and deuteron, respectively. The isotope shift, expressed as a velocity shift, is given by: $v_{\rm shift}=c(\lambda_{D}-\lambda_{H})/\lambda_{H}\simeq-81.6~{\rm km~s}^{-1}$. So, if we want to detect both the deuterium and hydrogen transitions, we require that the environment must not contain Doppler motions that exceed this value. Environments that are more turbulent than the isotope shift will cause the deuterium and hydrogen lines to be blended. This is made more difficult because we expect there to be just $\sim25$ deuterons for every one million protons (assuming the Standard Model of particle physics and cosmology).
%This can be achieved by observing a near-pristine environment with a spectrograph that measures the flux density of photons as a function of wavelength. Using this spectrum, we can then identify the absorption lines corresponding to the hydrogen and deuterium atomic transitions. 

\begin{figure}
    \centering
    \includegraphics[width=\columnwidth]{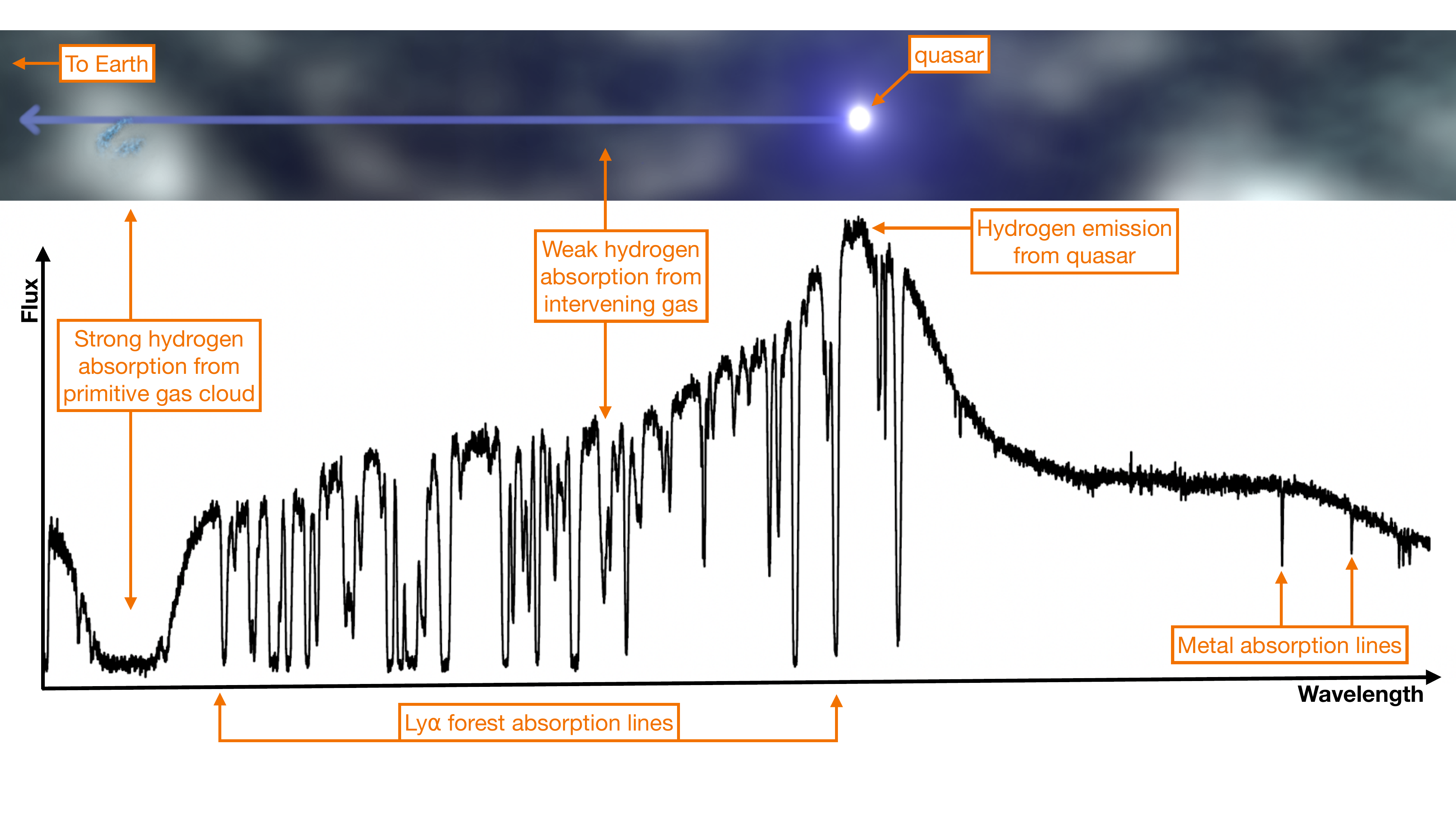}
    \caption{The top image shows a quasar with its emission directed towards our telescopes on Earth. The space in between is filled with galaxies and intergalactic gas that absorb the quasar light at wavelengths corresponding to atomic transitions (see the corresponding spectrum in the bottom panel). The expansion of the Universe causes these transitions to redshift relative to each other, leading to a forest of absorption lines at wavelengths less than the hydrogen \Lya\ emission of the quasar. Most of these absorption lines are due to weak hydrogen absorption. By pure chance, sometimes a large overdensity of primitive gas is intersected. This primitive gas cloud imprints the Lyman series absorption lines of deuterium and hydrogen on the quasar spectrum, and we can use these absorption lines to count the number of deuterium and hydrogen atoms that are in the gas cloud. In addition to deuterium and hydrogen, some heavy elements (i.e. `metals' made by stars, such as carbon and oxygen) absorb the quasar light, allowing us to assess the pollution of this gas since BBN.}
    \label{fig:qsoabs}
\end{figure}

It was first recognized by \citet{Adams1976} that there might be clouds of gas in the high redshift Universe that have very low Doppler motions, such that absorption by deuterium and hydrogen transitions could be imprinted on the light of a bright, unrelated background source. The idea is similar to looking at a lighthouse on a clear or foggy night. On a clear night, we can see a bright light from the lighthouse. On the other hand, when it is foggy, the fog makes the lighthouse appear dimmer. The brightest cosmic ``lighthouses'' in the Universe are quasars, which are rapidly accreting supermassive black holes in the centers of galaxies. By obtaining a spectrum of a quasar, we can study the clouds of gas that are between our telescopes and the  distant quasars. Furthermore, because the Universe is expanding, the absorption lines of each gas cloud are redshifted by different amounts, leading to a plentiful forest of absorption lines. This is known as the \Lya\ forest; an example of this approach is shown in Figure~\ref{fig:qsoabs}), where the absorption lines are almost entirely due to hydrogen atoms that are along the line-of-sight to the quasar.

\begin{figure}
    \centering
    \includegraphics[width=16cm]{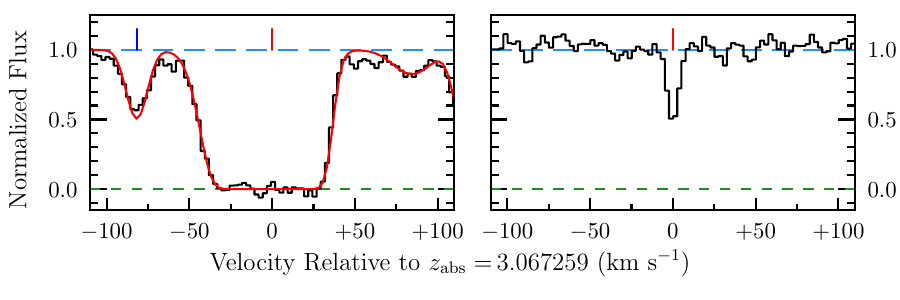}
    \caption{\emph{Left panel:} The black data shows an example of a deuterium and hydrogen absorption line system, due to hydrogen and deuterium Lyman-14 (Ly14); this is the transition from principal quantum number $n=1$ to $n=15$. The deuterium absorption (indicated by the blue tick mark above the spectrum) is located at $-81.6~{\rm km~s}^{-1}$ relative to the hydrogen absorption line (indicated by the red tick mark above the spectrum). The red curve is a model fit to the data (black histogram). Note that the hydrogen absorption is much stronger than the deuterium absorption. In order to determine the ratio of deuterium and hydrogen, we need to measure the relative strengths of these absorption lines. \emph{Right panel:} A silicon absorption line that is located at the same redshift as the deuterium and hydrogen absorption lines. Metals are chiefly produced by stars, and allow us to assess how pristine the environments are where deuterium and hydrogen can be measured.}
    \label{fig:DHabs}
\end{figure}

In some quasar spectra, there could be as many as several hundred hydrogen absorption lines, and all of these gas clouds contain deuterium atoms in them as well. However, most of these gas clouds are unsuitable to measure the ratio of deuterium and hydrogen atoms, because either:
(1) the deuterium absorption lines are too weak;
(2) the deuterium absorption lines are blended with other unrelated absorption lines; or
(3) the clouds have Doppler motions that exceed the isotope shift of $\sim-81.6~{\rm km~s}^{-1}$.
While this technique (called ``quasar absorption line spectroscopy'') was very promising, it took more than two decades of research and the advent of the 10 meter diameter Keck telescopes to find the first near-pristine (i.e. close to primordial) gas clouds where deuterium and hydrogen atomic transitions could be measured. An example of the deuterium and hydrogen absorption lines detected in a gas cloud using this technique is shown in the left panel of Figure~\ref{fig:DHabs}. In this example, the transition from principal quantum number $n=1$ to $n=15$ is shown (Ly14), which is just one absorption line of the Lyman series of hydrogen and deuterium). These same gas clouds also contain absorption lines from heavy elements (see the right panel of Figure~\ref{fig:DHabs} for an example). This tells us that most intergalactic gas clouds are not completely pristine reservoirs that have been untouched since the Big Bang; instead, their chemistry has been slightly altered by enrichment from stars. By measuring how many metals have been made relative to hydrogen in these environments, we can estimate how contaminated these environments are.

\begin{figure}
    \centering
    \includegraphics[width=16cm]{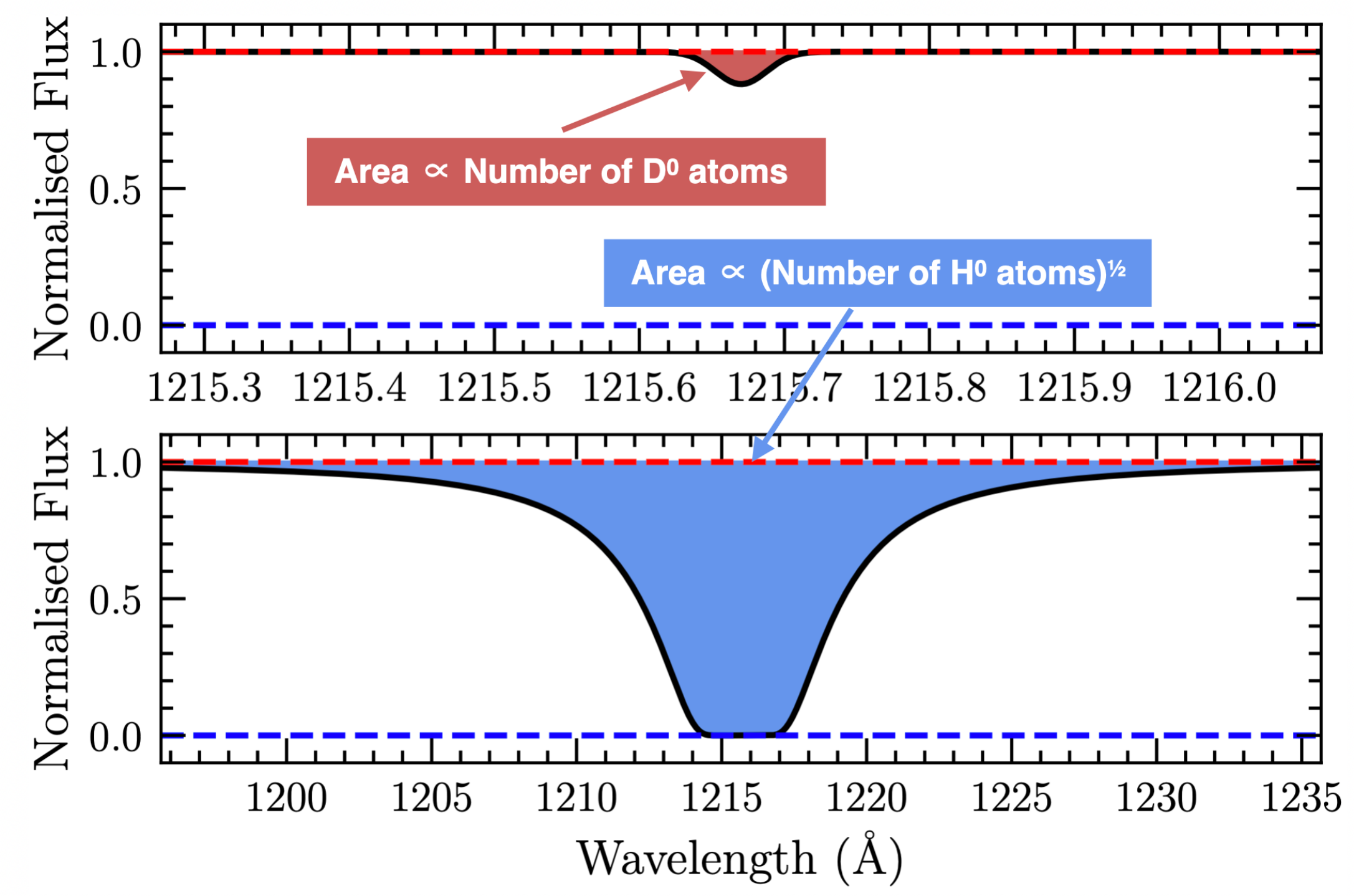}
    \caption{\emph{Top panel:} The equivalent width (i.e. the red shaded region that indicates the area under the continuum) of weak absorption lines is directly proportional to the total number of atoms that produce the absorption line. \emph{Bottom panel:} The equivalent width (blue shaded region) of very strong absorption lines (those with Lorentzian damped wings) is proportional to the square-root of the total number of atoms that produce the absorption line. Note the different x-axis scale of the two panels.}
    \label{fig:DHreliable}
\end{figure}

\begin{figure}
    \centering
    \includegraphics[width=16cm]{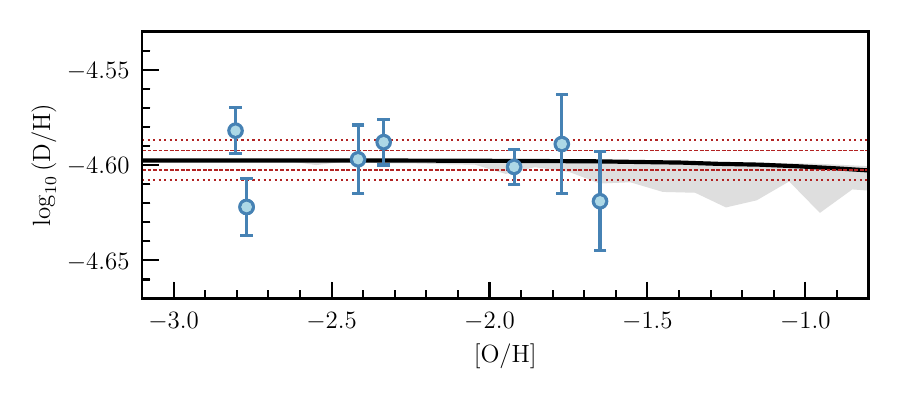}
    \caption{Seven independent measures of the deuterium abundance (i.e. the relative number of deuterium to hydrogen atoms) of gas clouds (blue symbols with error bars; \citealt{Cooke2018}). The x-axis shows the oxygen abundance, displayed on a log scale relative to solar (i.e. $-2$ is one-hundredth of the number of oxygen atoms in the sun relative to hydrogen, while $-3$ represents one-thousandth of the number of oxygen atoms). All seven measures are consistent with each other. The red horizontal lines represent 68 and 95 per cent confidence interval of the weighted mean value of these seven measures. The black line shows an example model of the chemical evolution of deuterium \citep{vandeVoort2018}, where the $3\sigma$ region is shown by the light grey band. Note that near-pristine systems have a deuterium abundance that is consistent with the primordial value to within $<0.1$ percent, based on this calculation.}
    \label{fig:DHmeasures}
\end{figure}

Since deuterium absorption lines were first detected, the technique used to identify these systems has been refined. The most reliable measurements use the rarest gas clouds currently known; they are among the most metal-poor gas clouds known anywhere in the Universe, and they are also mostly neutral gas clouds where the chemistry can be robustly measured. There are two key reasons why these systems are near-ideal environments: (1) The mostly neutral state of the gas gives rise to strong neutral hydrogen \Lya\ absorption. The wings of this absorption profile are Lorentzian damped (see the bottom panel of Figure~\ref{fig:DHreliable}), and the strength of this damped absorption is proportional to the (square root of the) total number of neutral hydrogen atoms along the line-of-sight. (2) The large number of neutral hydrogen atoms gives rise to many weak higher order Lyman series deuterium absorption lines. These weak deuterium absorption lines (see the top panel of Figure~\ref{fig:DHreliable}) are directly proportional to the total number of neutral deuterium atoms along the line-of-sight. Finally, given the similar ionisation potentials of deuterium and hydrogen (recall, the energy levels are almost identical), the ratios D\,\textsc{i}\,/\,H\,\textsc{i}\,$=$\,D\,/\,H are equivalent to within $0.1\%$ due to charge exchange reactions \citep{Cooke2016}.

Despite considerable work since the year 2000, just a handful of gas clouds are known where the deuterium abundance can be measured, demonstrating the rarity of the systems and how challenging this measurement is (a list of all deuterium absorbers, including references, is provided in the Further Information section). 
To date, the deuterium abundance has only been consistently measured for seven absorbers; these seven measurements are shown as the blue symbols in Figure~\ref{fig:DHmeasures}, as a function of the amount of stellar contamination, which is parameterized by the oxygen abundance, [O/H].\footnote{The square bracket notation used here is the logarithmic number abundance ratio of two elements, relative to the solar value, ${\rm [O/H]}={\rm log}_{10}({\rm O/H})-{\rm log}_{10}({\rm O/H})_{\odot}$. Therefore, ${\rm [O/H]}=0$ corresponds to the solar value, ${\rm [O/H]}=-1$ corresponds to one-tenth of the solar abundance, and so forth.} Even though these seven independent gas clouds have had a different amount of stellar processing, they are all statistically consistent with each other. This tells us that these gas clouds have retained a primordial relative composition of deuterium and hydrogen, and their ratio is the same as the value set just moments after the Big Bang.

Ideally, we would like to find completely pristine gas clouds that exhibit clean deuterium absorption lines (and some do exist, e.g. \citealt{Fumagalli2011}), however, such systems are exceedingly rare. Thankfully, the chemical evolution of deuterium is very straightforward; when a generation of stars is formed, all deuterium is burned by the star during its earliest phases (a process called astration). To illustrate this, imagine that a gas cloud of $10^6\,{\rm M}_{\odot}$ forms $10^3\,{\rm M}_{\odot}$ of stars. If we make the (simplistic) assumption that these stars completely burn deuterium and give back their birth amount of hydrogen, then the relative number of deuterium and hydrogen atoms would change by about 1 part in $\sim10^3$ (i.e. $\sim0.1\%$). Detailed models of chemical evolution, such as that shown in Figure~\ref{fig:DHmeasures} as the black solid line, have shown that the primordial deuterium-to-hydrogen ratio (D/H) is virtually unchanged, when the metals comprise $\lesssim1\%$ of the metals in the Sun \citep{vandeVoort2018}.

BBN calculations indicate that the relative abundance of deuterium and hydrogen atoms is highly sensitive to the expansion rate of the Universe and the density of baryons (i.e. ordinary matter; see Figure~\ref{fig:bbn_theory_obs}). If we assume the Standard Model of particle physics and cosmology, then the expansion rate of the Universe is fixed. Under this assumption, we can use the deuterium abundance to estimate the universal density of baryons. The baryon density is also a fundamental quantity that can be measured from the cosmic microwave background fluctuations. Quite amazingly, the baryon density that is inferred from the seven measures of D/H agrees with the baryon density derived from the cosmic microwave background at one percent precision! The incredible agreement of these two baryon density measures, based on completely independent physics and based on two epochs of the Universe separated by almost 400,000 years, represents a triumph to our cosmological understanding, and is one of the strongest confirmations of the standard cosmological model.

\subsection{Helium-4 --- The search for physics beyond the Standard Model}
\label{sec:helium}
As discussed in Section~\ref{sec:bbntheory}, \hef\ is the most abundantly produced primordial nuclide (with the exception of protons). Calculations of BBN indicate that the \hef\ abundance primarily depends on the expansion rate of the Universe. For historical reasons, both BBN calculations and observations usually report the primordial mass fraction of helium, unlike all other primordial element abundances that are quoted as number fractions. This difference is based on historical measurements of the Sun, which was found to contain $\sim25$\% helium by mass. At the time, there were no astrophysical sources known that could produce the large quantities of helium that were inferred in the Sun. In fact, this was one of the first triumphs of BBN calculations, which were able to postdict the mass fraction of \hef, giving extra credibility to the idea of a hot Big Bang. In modern times, it actually makes more sense to report the helium number abundance ratio, since:
(i) current observational probes intrinsically measure a number abundance ratio;
(ii) BBN calculations compute a ratio of volume densities, then calculate mass fractions as a post-processing step;
(iii) for non-primordial environments, the mass fraction depends on the fraction of mass in the form of metals.
There is still a tendency in the literature to report mass fractions, and it is common to see both mass fractions and number fractions reported. The convention is that primordial helium mass fractions are reported as $Y_{\rm P}$ (uppercase `Y'), while number abundances are reported as $y_{\rm P}$ (lowercase `y'). There are currently three approaches to measure the primordial abundance of \hef: (1) spectroscopic observations of metal-poor star-forming dwarf galaxies; (2) quasar absorption lines; and (3) the damping tail of the CMB power spectrum. These approaches will now be discussed in turn.

\begin{figure}
    \centering
    \includegraphics[width=16cm]{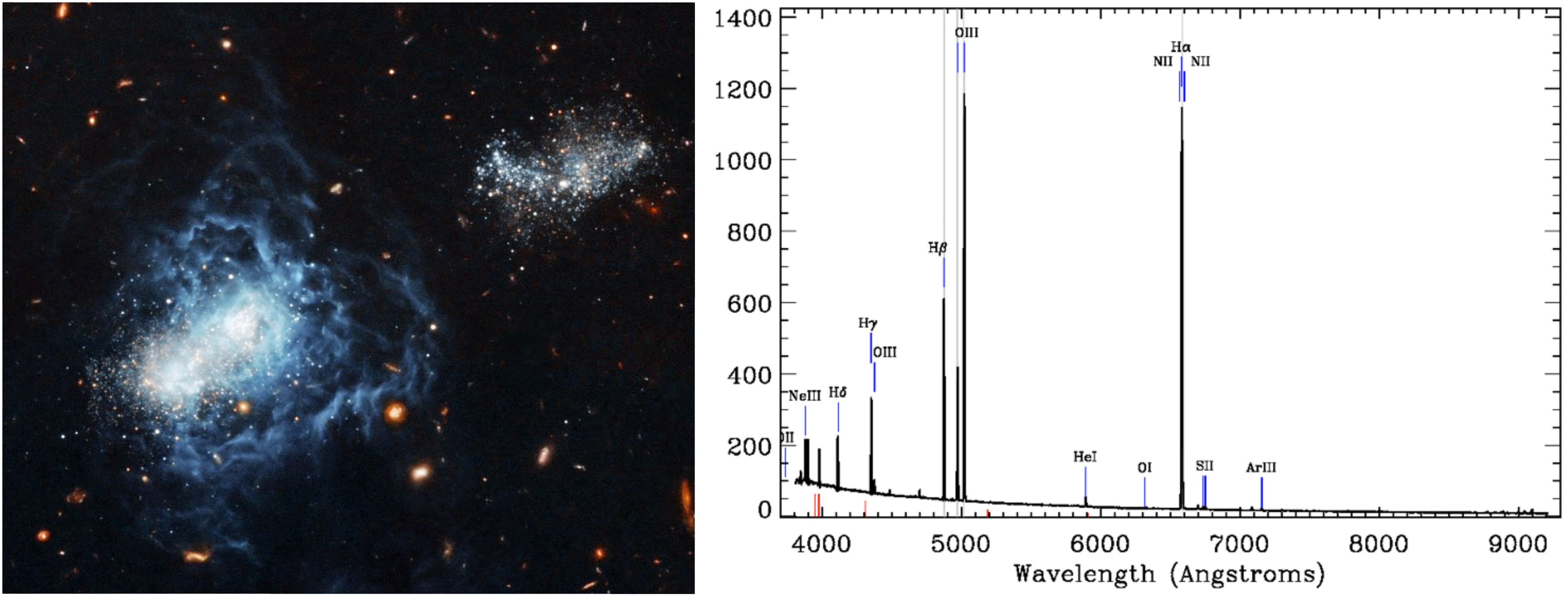}
    \caption{An image of I Zwicky 18 (left panel; credit: NASA, ESA, and A. Aloisi) and an optical spectrum (right panel; credit: SDSS). Note the blue appearance of the galaxy image and the emission lines that are detected in the spectrum, indicating that hot stars of O and B spectral type are ionizing the gas in this galaxy. The relative strengths of the emission lines can tell us about the physical and chemical properties of the excited gas.}
    \label{fig:IZw18}
\end{figure}

Galaxies that are currently forming a new generation of stars are referred to as `star-forming galaxies', and usually have blue colors owing to the presence of massive (and therefore short-lived) stars of spectral type O and B. These hot stars produce significant quantities of photons that are capable of ionising the atoms in the surrounding volume. Electrons that have been ionized from atoms eventually recombine with another atom and produce emission lines as the electron cascades down the atomic energy levels. An image and a spectrum of one of the most metal-poor star-forming galaxies currently known is shown in Figure~\ref{fig:IZw18}; this galaxy is called I Zwicky 18 (where the `I' is the roman numeral for the number one, and is pronounced `one Zwicky eighteen'). The ionized gas surrounding the O and B stars is generally referred to as an \HII\ region, because almost all of the hydrogen atoms have had their electron ionized. Similarly, `\HI' refers to regions that are mostly neutral, whereby most of the protons have captured an electron.

The relative strengths of the emission lines emanating from \HII\ regions depend on the chemistry of the gas (i.e. the relative abundance of each chemical element) and the physical conditions of the gas (e.g. the density, temperature, ionisation fraction, etc.). Most of the emission lines that come from \HII\ regions have a different dependence on the physical conditions, and by combining the information from many lines simultaneously, it is possible to determine both the physical and chemical properties of the gas. Measurements of the \hef/H ratio in different galaxies indicate that the \hef\ abundance gradually increases as stars produce heavy elements (see the right panel of Figure~\ref{fig:heliummetals} for an example). Therefore, the helium abundance that we measure in these star-forming galaxies does not reflect the primordial composition. Instead, we need to measure the helium abundance of many galaxies covering a range of metallicity, fit a linear relation to the helium and metal abundances, and extrapolate this linear fit to zero metallicity (see the blue linear relationship in the right panel of Figure~\ref{fig:heliummetals}).

Each individual \HII\ region needs to be observed, carefully analysed, and modelled in order to obtain a measurement of the helium abundance ($y$) of that \HII\ region. Most of the currently available data of \HII\ regions were acquired with longslit observations of metal-poor star-forming galaxies. The data are flux calibrated using observations of a flux standard star. The absolute flux does not matter; it is the \emph{relative} flux calibration between any two emission lines that matters most. Note that the current generation of flux standard stars are generally precise to within $\sim1-2\%$ at any two given wavelengths. Although, it is noteworthy that this relative flux calibration is generally worse than this near the H Balmer emission lines and \HeI\ emission lines due to features in the flux standard stars.

The spectrum of an \HII\ regions consists of a series of sharp emission lines superimposed on a continuum (see the right panel of Figure~\ref{fig:IZw18}). In the vicinity of each emission line of interest, the continuum is estimated by fitting a smooth function (usually a low order polynomial) to regions that are not affected by emission lines. The total emission line flux (and its associated error) is then taken to be the total (i.e. summed) flux above the fitted continuum level. Most \HII\ region studies attempt to model the flux of each emission line relative to the H$\beta$ emission line (the hydrogen transition from principal quantum number $4\to2$). This choice is somewhat arbitrary, but has become the standard in the field. Due to a series of physical processes, the measured flux ratio of a line relative to H$\beta$ is different from the intrinsic emission line ratio. Furthermore, each emission line depends on the physical conditions of the gas. There are several approaches that have been devised to extract information about the helium abundance from a set of emission line fluxes, and in what follows, we will consider one example calculation that allows us to separate the various processes that affect the emission line flux ratios. The equations used to model the hydrogen and helium emission lines are similar:

\vspace{-1.8cm}

\begin{align}
%%%%%%%%%%%%%%%%%%%%%%%
    \textrm{Helium emission lines:}\qquad
    \underbrace{%
    \vphantom{\scalebox{10}{(~)}}
    \frac{F(\lambda)}{F({\rm H}\beta)}}_\text{flux ratio}
    &=
    \underbrace{
    \vphantom{\scalebox{10}{(~)}}
    \frac{n({\rm He}^{+})}{n({\rm H}^{+})}}_\text{He$^{+}$ abundance}
    \times
    \underbrace{
    \vphantom{\scalebox{10}{(~)}}
    \frac{E(\lambda)}{E({\rm H}\beta)}}_\text{emissivity}
    \times
    \underbrace{
    \vphantom{\scalebox{10}{(~)}}
    \frac{1+a_{\rm H}({\rm H}\beta)/EW({\rm H}\beta)}{1+a_{\rm He}(\lambda)/EW(\lambda)}}_\text{stellar absorption}
    \times
    \underbrace{
    \vphantom{\scalebox{10}{(~)}}
    \frac{1+CR(\lambda)}{1+CR({\rm H}\beta)}}_\text{collisional correction}
    \times
    \underbrace{%
    \vphantom{\scalebox{10}{(~)}}
    10^{-f(\lambda)\,c({\rm H}\beta)}}_\text{reddening}
    ~~\times
    \underbrace{%
    \vphantom{\scalebox{10}{(~)}}
    f_{\tau}(\lambda)}_\text{optical depth}
    \\[-4pt]
%%%%%%%%%%%%%%%%%%%%%%%
    \textrm{Hydrogen emission lines:}\qquad
    \overbrace{
    \vphantom{\scalebox{3.5}{(~)}}
    \frac{F(\lambda)}{F({\rm H}\beta)}}_{~}
    &=
    ~~~~~~~~~~~~~~~~~~~~~
    \overbrace{
    \vphantom{\scalebox{3.5}{(~)}}
    \frac{E(\lambda)}{E({\rm H}\beta)}}_\text{~}
    \times
    \overbrace{
    \vphantom{\scalebox{3.5}{(~)}}
    \frac{1+a_{\rm H}({\rm H}\beta)/EW({\rm H}\beta)}{1+a_{\rm H}(\lambda)/EW(\lambda)}}_\text{~}
    \times
    \overbrace{
    \vphantom{\scalebox{3.5}{(~)}}
    \frac{1+CR(\lambda)}{1+CR({\rm H}\beta)}}_\text{~~~~~~~~~~~~~~~~~~~~~~~~~~~~~~~~~}
    \times
    \overbrace{
    \vphantom{\scalebox{3.5}{(~)}}
    10^{-f(\lambda)\,c({\rm H}\beta)}}_\text{~}
\end{align}

\noindent
where the left-hand side of the equation is the predicted flux ratio of a given emission line (with wavelength $\lambda$) relative to ${\rm H}\beta$. The number abundance of He$^{+}$ atoms relative to that of H$^{+}$, which is usually denoted $y^{+} \coloneqq n({\rm He}^{+})/n({\rm H}^{+})$, is a free parameter to be determined empirically and is one of the key parameters of interest to determine the helium abundance. The emissivities of each line, $E(\lambda)$, are based on detailed theoretical calculations that (usually) assume Case B recombination. Case B recombination assumes that all recombinations to the ground state produce a photon that excites a nearby atom. In the optically thick limit, transitions to the ground state never escape the cloud, and that energy is eventually radiated away through higher order transitions (e.g. such as the Balmer emission lines in the case of hydrogen). These Case B emissivities depend on density and temperature, and this dependence is different for each emission line. The stellar absorption term accounts for absorption lines that are intrinsic to the stellar continuum. This correction is smaller for emission lines that are much stronger than the continuum level. The collisional correction term accounts for hydrogen and helium atoms that are collisionally excited to higher energy levels. The excitations lead to recombinations that are not captured by the emissivities, which assume Case B recombination. The collisional contribution to the emission depends on the volume density of electrons, and the volume density of neutral hydrogen atoms. The reddening term accounts for wavelength dependent absorption due to dust along the full line-of-sight. The variable $f(\lambda)$ is the reddening law, while the variable $c({\rm H}\beta)$ is a free parameter to be determined from the data. The optical depth term accounts for photons that are emitted but subsequently reabsorbed or scattered out of our line-of-sight. This correction is stronger for the triplet emission lines of neutral helium (i.e. transitions of helium atoms where the spins of the two electrons are aligned), compared to the singlet emission lines of neutral helium (where the electron spins are anti-aligned). Finally, several of these terms depend on the electron density and the electron temperature of the gas, and these are considered free parameters that are determined empirically from the data.

The model parameters are determined so that all hydrogen and helium emission lines are simultaneously reproduced. The total helium abundance of a galaxy is then given by the sum of all ionisation states of helium, $y=y^{+} + y^{++}\equiv [n({\rm He}^{+}) + n({\rm He}^{++})]/n({\rm H}^{+})$, where the $y^{++}$ abundance is typically a $\sim1\%$ correction. The metallicity of the galaxy is also measured using emission lines (usually oxygen or sulphur). A linear function is fit to a sample of $y$ and metallicity values (e.g. O/H; see Figure~\ref{fig:heliummetals} for an example). The primordial helium abundance is then found by extrapolating this linear fit to zero metallicity; the intercept of this linear fit corresponds to the primordial value. Current determinations of the primordial helium abundance using this approach are limited by systematic uncertainties at the $\sim1-2$ percent level. While it is currently unclear what causes this systematic limit, the emission lines of many galaxies are not well-modelled by the currently adopted procedures. This could point to problems with the data collection, or one or more of the aspects of the modelling procedure.

Another approach that has been used to infer the primordial helium abundance uses quasar absorption line spectroscopy (see Section~\ref{sec:deuterium} for an explanation of this technique) of metal-poor gas clouds at high redshift ($z\sim1-2$). This approach is qualitatively similar to the approach used to measure the deuterium abundance, however, there is a key difference between these approaches. For deuterium, one tries to find mostly neutral gas clouds to facilitate the detection of many deuterium Lyman series transitions. However, these mostly neutral gas clouds absorb essentially all of the background quasar photons with energies greater than the ionisation potential of hydrogen ($13.6~\rm{eV}\equiv912$\AA). Since all of the ground-state helium transitions occur at wavelengths $<912$\AA, there is insufficient quasar flux to detect both the helium and hydrogen absorption lines in predominantly neutral gas clouds. Therefore, in order to detect both helium and hydrogen absorption lines, the gas cloud needs to be transparent to photons at wavelengths $<912$\AA; this occurs for gas clouds that are mostly ionized, with a column density of neutral hydrogen $N({\rm H}\,\textsc{i})\lesssim10^{17}~{\rm atoms~cm}^{-2}$. The helium abundance has only been measured in one gas cloud with a metallicity of $\sim1/30$ solar, which is a similar metallicity to the most metal-poor star-forming galaxies used to measure the helium abundance. The measurement precision of this technique is currently at the $\sim10$ percent level. Further improvements to this measurement will require new suitable gas clouds to be identified and a better understanding of the ionisation properties of the gas cloud. Thus, for the foreseeable future, this approach is not as competitive as using \HII\ regions in the context of measuring the primordial helium abundance.

An alternative and very promising approach to measure the primordial helium abundance utilizes the power spectrum of temperature fluctuations imprinted on the Cosmic Microwave Background radiation. This approach has the advantage that helium is purely primordial during recombination, because there are no channels (i.e. stars) to produce/destroy \hef\ during the first 400,000 years after BBN. The key physical process that is sensitive to the helium abundance is diffusion damping. Note also that if more \hef\ is produced during BBN, that means less hydrogen is made, and vice-versa. Diffusion damping (sometimes referred to as Silk damping) is a process whereby the temperature fluctuations are smoothed out on small angular scales (i.e. high multipoles of the power spectrum). The damping is a result of photons being constantly scattered by charged particles (mostly electrons) prior to recombination. Photon scattering stops when the temperature of the Universe drops to the point that electrons can recombine with nuclei, at which point photons can free stream. Electrons recombine with helium first, followed later by hydrogen recombination; this is because helium has a higher ionization potential than
\begin{wrapfigure}{r}{0.5\textwidth}
    \centering
    \includegraphics[width=8cm]{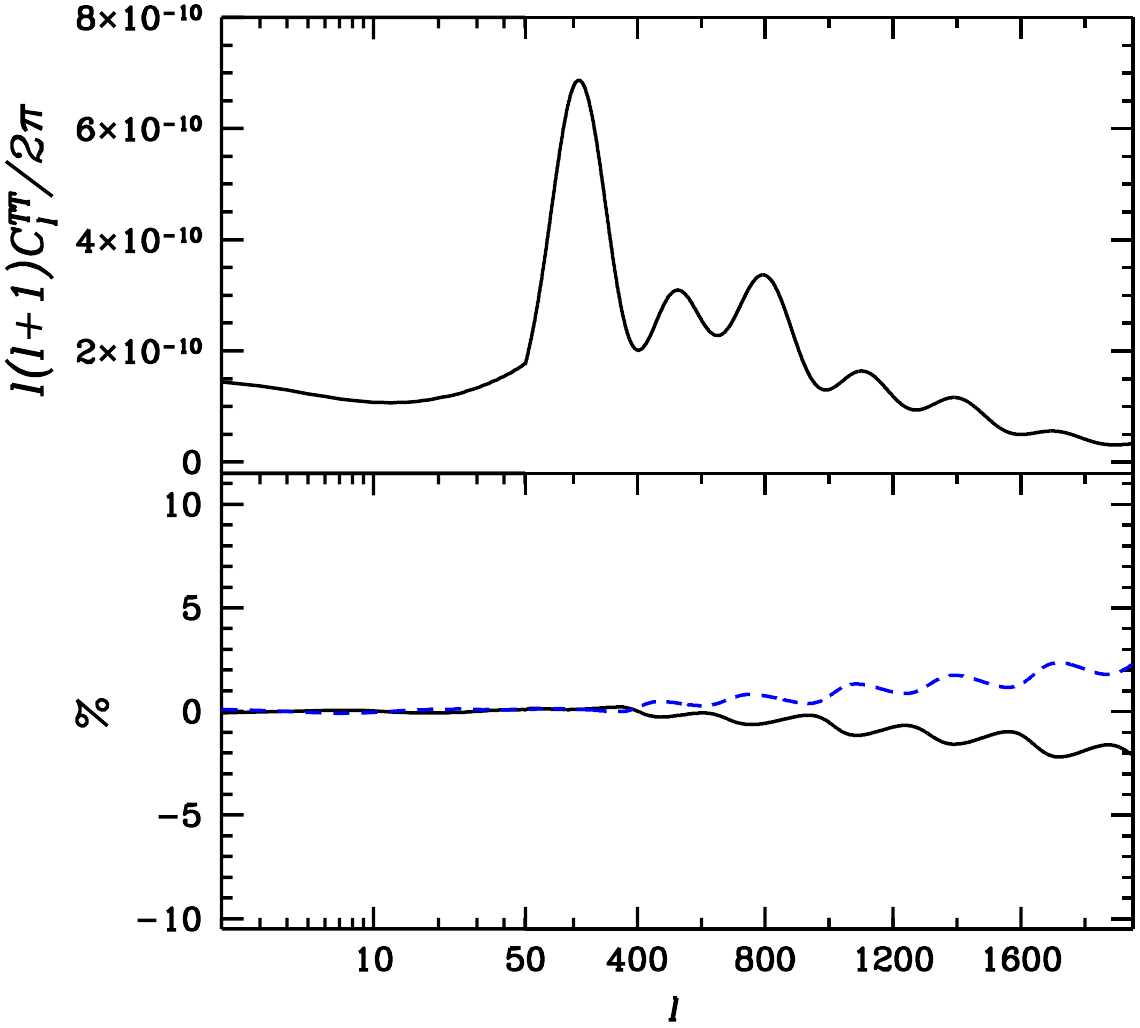}
    \caption{The power spectrum of the Cosmic Microwave Background radiation (top panel), and the deviations of the power spectrum (in percent) due to changes of the helium abundance (bottom panel). The blue dashed and solid black lines represent a decrease and increase of the primordial helium abundance by 10 per cent. Reproduced from \citet{TrottaHansen2004}.}
    \label{fig:silkdamping}
\end{wrapfigure}
hydrogen (i.e. recombined electrons are more difficult to ionize from helium than hydrogen). If more helium is made during BBN, there are fewer electrons between helium and hydrogen recombination, leading to less scattering. This allows photons to free stream further, damping the perturbations and reducing the CMB power at small angular scales. The effect is relatively small (see Figure~\ref{fig:silkdamping}), and the current inference on the primordial helium abundance using this approach is $\sim10$ percent. CMB experiments that are planned for the early 2030s (called `CMB stage 4' experiments) will target small angular scales; one of the aims is to measure the primordial abundance of helium with a precision of $\sim2$ percent. This level of precision is a `best case' scenario, and is not higher precision than that currently delivered by \HII\ regions. Nevertheless, determining the primordial helium abundance from the CMB with a precision of a few percent will offer an important cross-check that will help to validate both approaches.

\subsection{Helium-3 --- Helpful but hard to find}
% Trials and Triumphs of Tralphium
% Tricky Tralphium

\begin{figure}
    \centering
    \includegraphics[width=16cm]{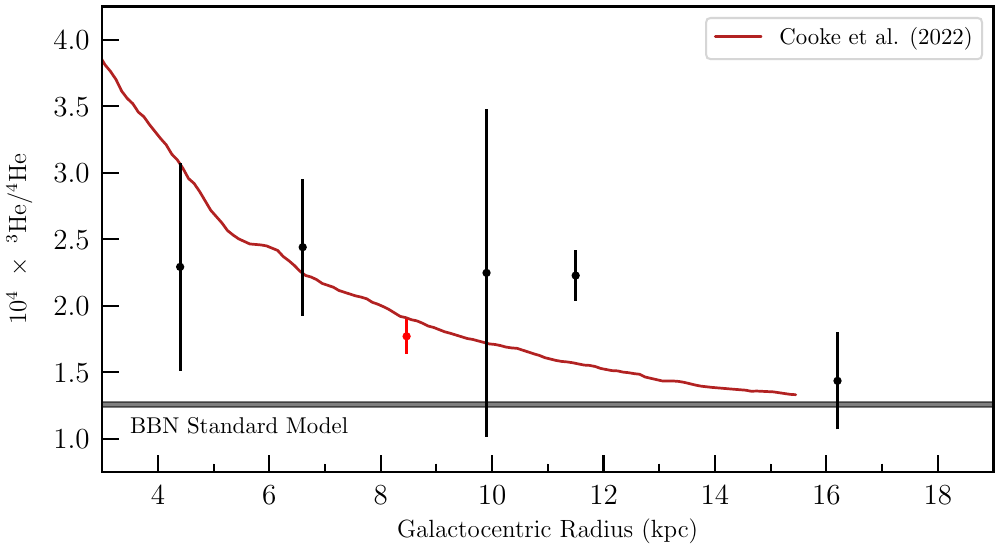}
    \caption{The present-day radial distribution of \het\ in the Milky Way (symbols with error bars) together with a detailed model of Galactic chemical evolution  (red curve; \citealt{Cooke2022}) and the primordial abundance (gray horizontal band labeled ``BBN Standard Model''). The black points correspond to measurements of the 8.7 GHz hyperfine structure line of \het$^{+}$, while the red symbol represents a measure of the \het/\hef\ isotope ratio of the Orion Nebula, using absorption line spectroscopy.}
    \label{fig:heliumRgal}
\end{figure}

For several reasons, \het\ has proven to be a very challenging primordial nuclide to find. First, almost all \het\ and \hef\ transitions are very close in wavelength; the difference is one neutron in the nucleus, and the isotope shift is typically $\lesssim10~{\rm km~s}^{-1}$ (see Section~\ref{sec:deuterium} for a discussion about isotope shifts). The proximity of these transitions, and the fact that \hef\ is 10,000 times more abundant than \het, has made measurements of the \het\ abundance nearly impossible. \het\ has been securely detected in the Milky Way, including a small handful of \HII\ regions and a few environments in our Solar System (Jupiter, in particular, provides the best current estimate of the pre-solar \het/\hef\ abundance). Unfortunately for measurements of the primordial abundances, the Milky Way has experienced a significant amount of chemical enrichment, and the \het\ abundances measured in the Solar System and in some of the Milky Way \HII\ regions do not reflect the primordial value. However, despite the significant build-up of metals in the Milky Way, the latest models of Galactic chemical evolution suggest that the abundance of \het\ in the outskirts of galaxies like the Milky Way may be close to primordial (see the red curve in Figure~\ref{fig:heliumRgal}).

Other than the direct measurement of \het/\hef\ in Jupiter's atmosphere, the first extra-solar measurement of the \het\ abundance was based on the 8.7~GHz spin-flip transition of \hetp\ (i.e. singly ionized \het). This transition occurs between the two hyperfine levels of the \het\ ground state, which is split due to the non-zero nuclear spin of the \het\ nucleus. Furthermore, the nuclear spin of \hef\ is zero, so the ground state of \hef\ does not exhibit hyperfine splitting. This makes it relatively straightforward to detect the \hetp\ emission line, since it does not need to be separated from any \hef\ emission. After many decades of work, there are just five reliable measurements of the \het\ abundance in Galactic \HII\ regions \citep{BalserBania2018}. The most distant, and well-characterized of these \HII\ regions is called `Sh2-209', and is 12.7~kpc from the Galactic centre, where models of Galactic chemical evolution predict a very modest enhancement in the amount of \het\ above the primordial value (see Figure~\ref{fig:heliumRgal}).\footnote{Note that this \HII\ region was previously thought to be at a galactocentric distance, $R_{\rm gal}\sim16~{\rm kpc}$ before the distance to the resident stars were measured recently with high precision using the \emph{Gaia} satellite. The methodology to estimate the \het/\hef\ value depends on distance (see text for further details). Therefore, `Sh2-209' is shown in Figure~\ref{fig:heliumRgal} at a distance of $\sim16~{\rm kpc}$.} The value measured in this \HII\ region agrees with the Standard Model value to within $\sim30$ per cent.

To calculate the \het/H abundance of these \HII\ regions, a model of the emitting gas is required, and an ionization correction needs to be applied to account for any neutral \het\ that is present in the \hetp\ zone. The models assume that the gas is optically thin, in local thermodynamic equilibrium, and consists of only H and He gas. Based on the measured brightness temperature of the continuum and the hydrogen recombination lines, the electron temperature can be estimated from the line-to-continuum ratio. Using radio recombination lines of H$^{+}$ and \hefp, a measure of \hefp/H$^{+}$ can be determined. The density structure of the emitting gas is modelled assuming that the \HII\ region is a homogeneous, spherically symmetric cloud. Given the electron temperature, the \hefp/H$^{+}$ value, and the distance of the \HII\ region from the Sun, the electron density and the diameter of the \HII\ region can be determined. This procedure is iterated until the H$^{+}$ and \hefp\ radio recombination emission lines are reproduced with the model. The \hetp/H$^{+}$ value is also adjusted to match the \hetp\ 8.7~GHz emission.

In the final step of the analysis, an ionization correction is applied to account for neutral \het\ that is coincident with the H$^{+}$ emitting region. This correction requires several assumptions. First, the ionization of \het\ and \hef\ are identical, such that $n(^{3}{\rm He}^{+})/n(^{3}{\rm He}) \equiv n(^{4}{\rm He}^{+})/n(^{4}{\rm He})$. This is a reasonable assumption, given the nearly identical ionization potentials of \het\ and \hef. Second, a simple model is constructed for the \hef/H abundance as a function of galactocentric radius. This model is based on a linear fit to the \hefp/H$^{+}$ ratios of two \HII\ regions (M17 and Sh2-206) that do not show evidence of neutral helium (i.e. in these regions, it is assumed that \hefp/H$^{+}={\rm He/H}$). This procedure has been applied to five Milky Way \HII\ regions covering a wide range of galactocentric radius (see black points in Figure~\ref{fig:heliumRgal}).

\begin{figure}
    \centering
    \includegraphics[width=16cm]{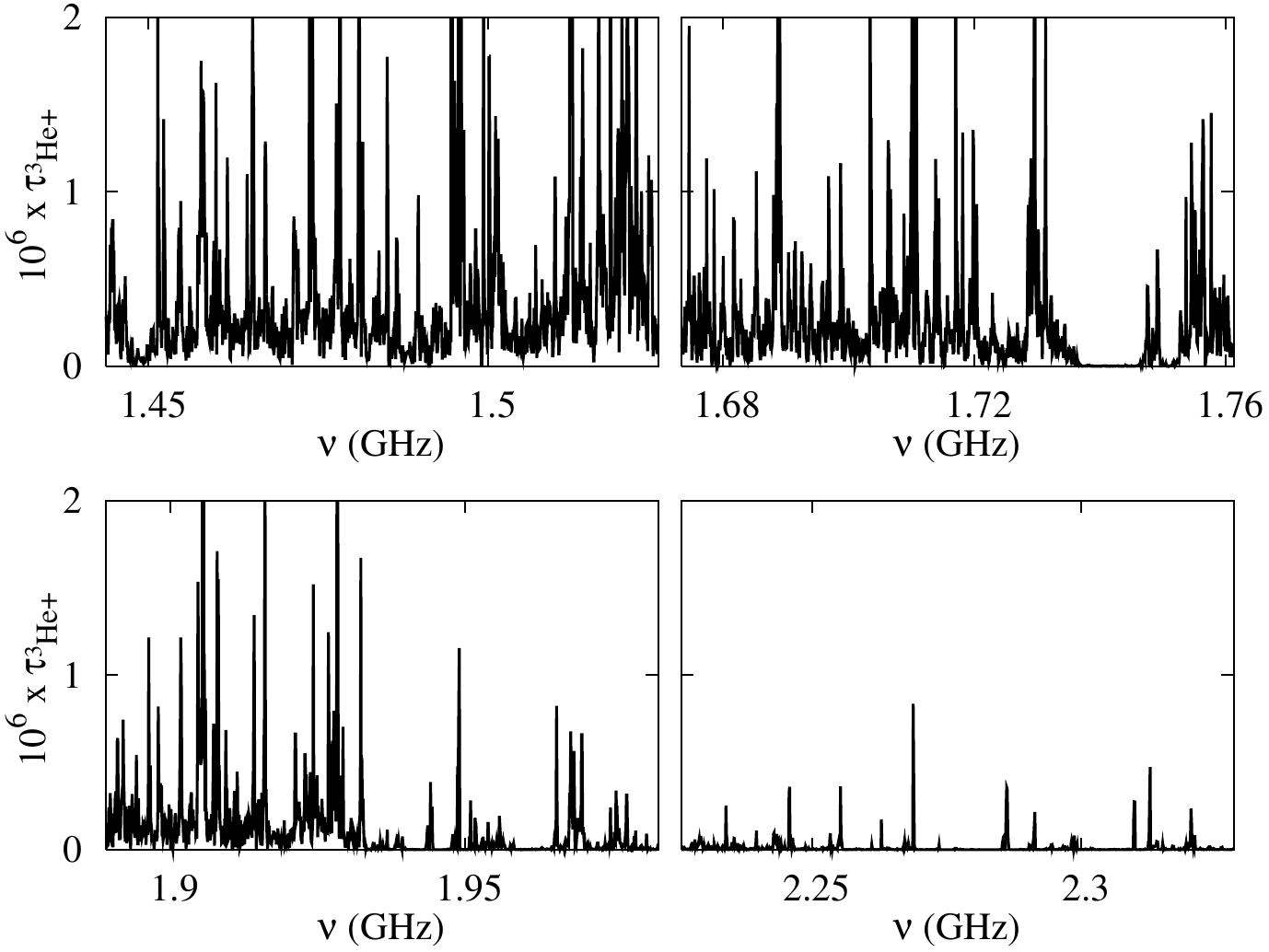}
    \caption{The optical depth of \hetp\ absorption ($\tau_{^{3}{\rm He}^{+}}$), that would be detected in absorption along the line-of-sight to a distant quasar at redshifts $z=5.0, 4.2, 3.6, 2.9$ (respectively, top-left, top-right, bottom-left, bottom-right). Reproduced from \citet{McQuinn2009}.}
    \label{fig:heliumIGM}
\end{figure}

\begin{figure}
    \centering
    \includegraphics[width=16cm]{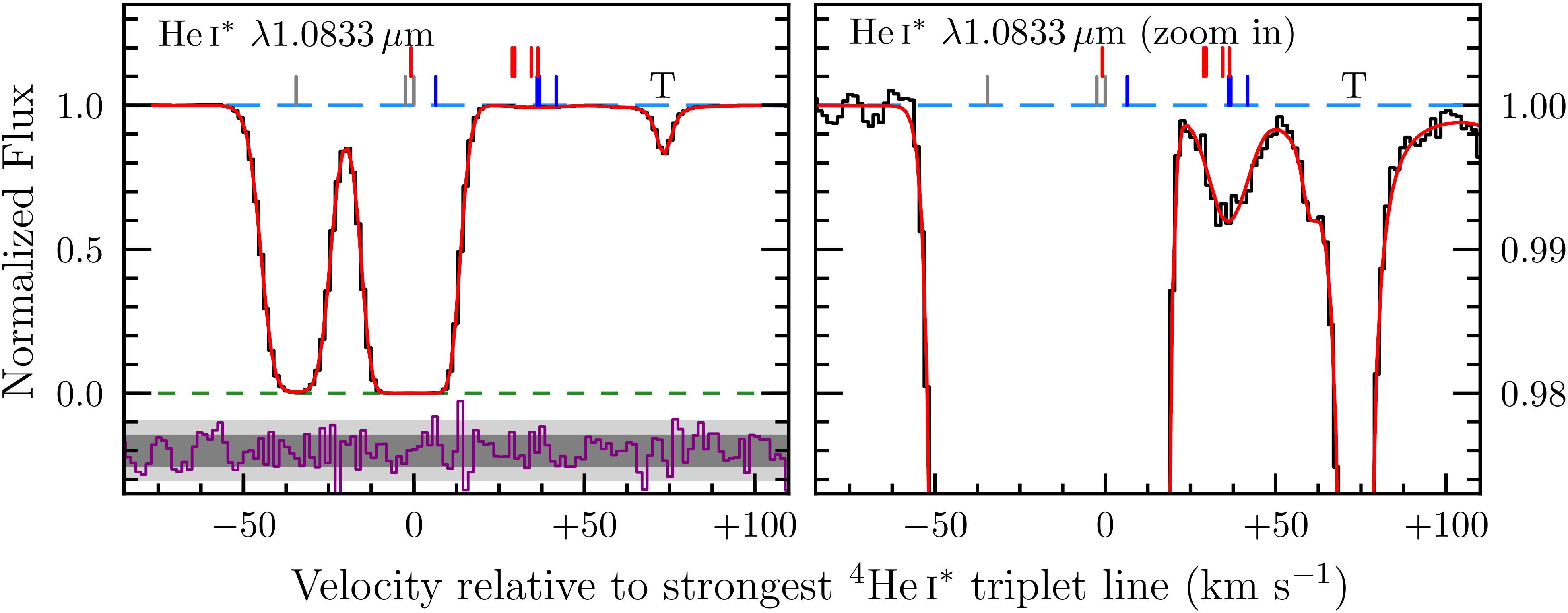}
    \caption{The left panel shows \hef\ absorption lines (three grey tick marks above the spectra indicate the transitions from the 2S to 2P state of triplet \hef. The blue and red tick marks indicate the corresponding absorption due to \het\ (note that there are many more \het\ absorption features than \hef\ absorption features, due to the hyperfine splitting of \het). The right panel shows a zoom-in of the left panel (note the different y-axis scale) to highlight the extremely weak \het\ absorption that is associated with the much stronger \hef\ absorption. The red line shows a model fit to the data; the purple histogram at the bottom of the panel represents the fit residuals, (data$-$model)/error, and the dark/light grey band represents the 68 and 95 perfect deviations, respectively. The absorption line marked with a `T' is due to an absorption line from the Earth's atmosphere. Modified version of the figure presented in \citet{Cooke2022}.}
    \label{fig:he_isotope}
\end{figure}

An alternative to this approach has been proposed that uses the same 8.7~GHz transition to detect \hetp\ in absorption against the light of a radio-bright background quasar close to helium reionization at redshift $z=3.5$ (see Figure~\ref{fig:heliumIGM}). While the astronomical telescope facilities do not currently exist to perform this measurement, this is an exciting goal for future facilities, because the intergalactic regions that would be probed with this technique are almost certainly of near-primordial composition, similar to those discussed in Section~\ref{sec:deuterium}. Furthermore, the measurement does not require the same level of modelling detail that is required of \HII\ regions.

More recently, it has been realized that the helium isotope ratio (\het/\hef) may provide a more robust determination of primordial \het\ production; given the similarity of the ionization potentials, there is no need to apply an ionization correction. However, this approach would require a measurement from the same transition, requiring the isotope shift to be resolved. The first example of this approach was demonstrated using absorption line spectroscopy of a star in Orion to study the gas that lies at the edge of the Orion Nebula (see Figure~\ref{fig:he_isotope}). This approach is qualitatively similar to the approach used to measure the deuterium abundance of gas clouds at high redshift (see Section~\ref{sec:deuterium} for a discussion about this approach). The helium isotope of this one measurement is shown in Figure~\ref{fig:heliumRgal} by the red point. Quite amazingly, this measurement suggests that post-BBN production of \het\ is relatively minimal, even in the Milky Way galaxy that is far from primordial. Future measurements using this technique may help to precisely pin down the Galactic chemical evolution of \het, and also obtain a determination of the primordial helium isotope ratio from the outskirts of the Milky Way. It is also possible to apply this technique to gas clouds at higher redshift, where the Universe has had less time to pollute the primordial signature.

\subsection{Lithium-7 --- The Cosmic Lithium Problem}
\label{sec:lithium}

\begin{figure}
    \centering
    \includegraphics[width=16cm]{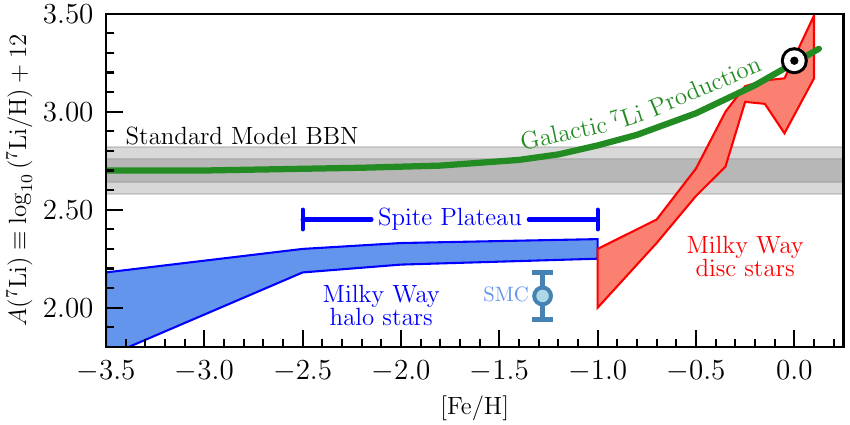}
    \caption{The chemical evolution of \lis\ in the Milky Way. The x-axis shows the iron abundance on a log scale relative to solar (i.e. $0$ is the abundance of iron in the Sun, $-1$ is one-tenth the value of the Sun, etc.). The primordial \lis\ abundance assuming the Standard Model is shown by the grey horizontal region, where the dark and light shades represent the 68 and 95 percent confidence regions. As the Universe becomes more enriched with iron, some \lis\ is made by stars and cosmic rays. The build-up of \lis\ as the iron abundance increases is shown by the green solid line, and the solar value is given by the $\odot$ symbol. The measured value in stars is shown by the blue and red shaded regions. The `Spite Plateau' refers to the roughly constant value of \lis/H between $-2.5<{\rm [Fe/H]<-1.0}$. At extremely low metallicity, ${\rm [Fe/H]}<-2.5$, the Spite Plateau declines. The \lis\ abundance of a sightline towards the Small Magellanic Cloud (SMC) (based on interstellar absorption line spectroscopy; see text) is shown as the light blue symbol with an error bar.}
    \label{fig:lithium}
\end{figure}

The best available observational determination of the primordial \lis\ abundance is based on measurements of the \lis\ absorption in the atmospheres of old metal-poor stars that orbit the Milky Way galaxy. This is not a straightforward measurement, because stars are quite good at burning \lis. Convective motions (particularly in cool stars) mix the \lis\ near the surface of the star into the deeper layers where \lis\ is burned. Meanwhile \lis-deficient material is brought to the surface. To mitigate the effects of this process, observations typically focus on the warmest metal-poor stars, which have thin convective zones, and the measured \lis\ abundance does not correlate with temperature. This key realization came from \citet{SpiteSpite1982}, who reported the first measurement of the \lis\ abundance in metal-poor stars. Specifically, using only the hottest halo stars, \citet{SpiteSpite1982} found that the \lis\ abundance of these stars is independent of temperature and metallicity, and has a very small scatter. This seemingly constant \lis\ abundance is referred to as the `Spite Plateau' (see Figure~\ref{fig:lithium}), and this value has remained impressively constant over the last few decades.

The process of measuring the \lis/H abundance of metal-poor stars is well-established. Suitable stars of spectral type F, G, or K need to be identified with thin convective zones and low metallicity. These stars are observed at high spectral resolution and high signal-to-noise ratio to confidently detect the Li\,\textsc{i}\,$\lambda6707$\,\AA\ stellar absorption lines against the continuum of the star.\footnote{To be clear, the Li\,\textsc{i} absorption lines described here are formed in the atmosphere of the star being observed, unlike the absorption lines discussed in the previous sections, that form in gas clouds that reside somewhere along the line-of-sight to a background light source.} In order to convert the observed absorption line into a measure of the \lis\ abundance, the physical properties of the star need to be determined. The most crucial parameter (in the context of lithium abundance measurements) is the effective temperature of the star; for every $\sim10$~K change to the effective temperature, the lithium abundance changes by $\sim0.01$~dex. There are several methods to measure the temperature, and these usually agree to $\lesssim200$~K. The surface gravity of the star can then be estimated by assuming ionization equilibrium between Fe\,\textsc{i} and Fe\,\textsc{ii} lines. The microturbulence of the model atmosphere is then estimated by ensuring the weak and strong Fe\,\textsc{i} absorption lines provide a consistent measure of the total iron abundance of the star.

For simplicity and computational efficiency, stellar atmospheres are usually modelled in one dimension (the radial coordinate) and the absorption lines are assumed to form in regions of local thermodynamic equilibrium (LTE). For the unevolved stars that comprise the Spite plateau, this is a reasonable assumption. The effects of non-LTE line formation and the granulation of the stellar atmosphere (i.e. `3D' effects) cause overall corrections of $<0.05~{\rm dex}$. Thus, the modelling procedure does not introduce large systematic uncertainties in the determination of stellar \lis\ abundances.

For many years, it was assumed that the Spite plateau was representative of the primordially produced abundance of \lis/H. However, at the turn of the millennium, the first Wilkinson Microwave Anisotropy Probe (WMAP) results provided a tight bound on the baryon density, and this revealed that the Spite plateau disagreed significantly with the Standard Model value, based on the CMB-derived baryon density. This problem has come to be known as the `Cosmic Lithium Problem' \citep{Fields2011}. Many groups have remeasured the lithium abundance with increasingly detailed stellar atmosphere models, and the result was essentially unchanged; actually, the improved models have made the disagreement with the Standard Model slightly more pronounced. Several nuclear physics groups remeasured some of the reaction rates involving \lis\ and $^{7}$Be, but the refined measurements of the reaction rates also made the disagreement even more pronounced. The current consensus is that one of the following possibilities must be the solution to the Cosmic Lithium Problem:

\smallskip

\begin{itemize}
    \item the stars that are used to infer the primordial lithium abundance may have destroyed some lithium during their lives
    \item there may be missing or poorly measured nuclear reaction rates, particularly those involving the destruction of $^{7}$Be
    \item there could be physics beyond the Standard Model that suppresses the production of $^{7}$Be, but does not appreciably affect the production of D, \hef, and \het.
\end{itemize}

\smallskip

Of these scenarios, the first possibility is perhaps the most likely; several groups have now measured the \lis\ abundance in extremely metal-poor stars, and found that the Spite Plateau appears to break down at metallicities less than 1/300 of solar (see Figure~\ref{fig:lithium}). Furthermore, by examining the lithium abundance of stars in metal-poor globular clusters (which are presumed to form from gas of the same initial chemical composition and at the same time), the \lis/H abundance is found to depend on effective temperature and stellar luminosity. These trends can be explained by stellar structure models that include atomic diffusion and turbulent mixing below the convective zone of the star \citep{Korn2007}. Overall, this gives further weight to the idea that the Cosmic Lithium Problem is a result of \lis\ astration.

In an attempt to remedy the problem of \lis\ burning in stars, a measurement of the interstellar abundance of \lis\ was made in the lowest metallicity environment where this measurement is currently possible --- the Small Magellanic Cloud (SMC; \citealt{Molaro2024}). The technique used is similar to quasar absorption line spectroscopy (see description in Section~\ref{sec:deuterium}), but instead of a quasar, a bright O star is used as the background light source to probe the interstellar medium of the host galaxy. Unfortunately, most of the lithium in this neutral gas cloud is singly ionized; transitions of Li\,\textsc{ii} occur at energies $>1\,{\rm Ryd}$, and are not observable because the significantly more abundant hydrogen atoms absorb all of the continuum photons above $1~{\rm Ryd}$. The only transitions available come from neutral \lis, and require a substantial ionization correction (about a factor of $\sim100\times$ the measured value) to convert the Li\,\textsc{i} column density into a measure of the total lithium column density along the line-of-sight. There are several ways to perform the ionization correction that will not be discussed in detail here, but it should be noted that all corrections give mostly similar values of the \lis/H ratio. However, it is worth bearing in mind that there is still some uncertainty in this correction. Incidentally, the determination of \lis/H in the SMC is consistent with the Spite Plateau, and inconsistent with the value expected from the Standard Model BBN (see Figure~\ref{fig:lithium}). Thus, further work is needed to elucidate the cosmic chemical evolution of \lis, and pin down with confidence an observational determination of the primordial \lis\ abundance.

\section{The future of Big Bang Nucleosynthesis}

Having summarized the current status of BBN, it is a worthwhile exercise to peer into the crystal ball, and consider what developments and goals should be the focus of future studies of BBN.

\subsection{BBN theory}

Numerical calculations of BBN are only as reliable as the input data, specifically the nuclear reaction rates, and the neutron mean lifetime. While it would be a wonderful achievement to fully calculate the products of BBN from first principles and theoretical calculations alone, this is quite unrealistic. For the foreseeable future, the input data are largely driven by experimental measures of the nuclear reaction rate cross-sections as a function of energy. There are some inherent uncertainties associated with these measurements; indeed, the precision with which several reaction rates have been measured is known to be the leading cause of the uncertainty of BBN theory. The reaction rates that are of highest current interest are those involving deuteron fusion:
\begin{eqnarray}
    d + d &\longrightarrow& ^{3}{\rm H} + p\\
    d + d &\longrightarrow& ^{3}{\rm He} + n
\end{eqnarray}
which primarily affects the primordial abundance of deuterium. The most important energy range to pin down these rates is between $0.05-0.4~{\rm MeV}$. The next most important measurement for BBN theory is the neutron mean lifetime, which primarily affects the synthesis of \hef. There are two experiments that are used to measure the neutron mean lifetime, known as the `beam' and the `bottle' --- both disagree with one another. The `bottle' approach counts the number of neutrons at various time intervals, while the `beam' experiments count the decay products of neutrons. Recent improvements to this measurement (particularly the bottle) have not helped to resolve the disagreement. For now, both approaches estimate the neutron mean lifetime with a precision of $\sim0.1\%$, but differ from each other at $\sim1\%$. Future high precision measurements of \hef\ will require this fundamental measurement to be pinned down.

\subsection{BBN observations}

Future primordial abundance measurements have the potential to strongly test the Standard Model of particle physics and cosmology. All of the current measures of the primordial abundances could be improved --- some more than others! Provided the uncertainties associated with BBN theory can also be improved (particularly the $d+d$ rates), the primordial abundance of deuterium has a lot of promise. CMB Stage 4 experiments (i.e. CMB experiments that are currently planned for the early 2030s) are poised to pin down the baryon density with higher precision than is likely possible with BBN. However, the combined power of CMB+BBN will provide the tightest constraints yet on departures from the Standard Model expansion rate of the Universe. Current samples of D/H measurements appear to be limited by statistics, not systematics, and so there is scope to continue improving the D/H measurement precision with the forthcoming generation of extremely large telescope facilities of 30+\,m aperture. First of all, larger telescopes will provide access to fainter (and considerably more numerous) background quasars and allow the statistics to be improved by a factor of $\gtrsim100$. Other future facilities aim to target D/H absorption line systems at lower redshift, where blending from unrelated absorption features in the \Lya\ forest are of less concern, and ever cleaner measurements are likely possible.

The next generation (Stage 4) CMB experiments, which are planned for the early 2030s, aim to pin down the primordial helium abundance with a precision of two percent; this is unlikely to compete with the precision that is possible from \HII\ regions. However, a lot more work is needed to overcome the systematic uncertainties that affect the current generation of measurements. One of the most exciting future measurements in cosmology is to detect the slight increase of the photon temperature just prior to BBN, due to the final interactions between the higher energy neutrinos and electron-positron annihilation. This will be a truly remarkable demonstration of our understanding of the early Universe. However, in order to detect this process with $2\sigma$ confidence, the primordial helium abundance would need to be measured with a precision of $0.1\%$, which is not inconceivable, but perhaps optimistic.

\het\ has proven to be a very challenging element to detect. Current observations of \het\ are limited to the galaxy, but there is a chance that this isotope may be detectable in absorption along the line-of-sight to some gamma-ray bursts. This perhaps offers our best opportunity yet of measuring the primordial helium isotope ratio, \het/\hef. However, this challenging goal will likely need to wait until the next generation of extremely large telescopes. Gamma-ray bursts have relatively short lifetimes and the detection of \het\ requires extremely high signal-to-noise ratio.

One has to hope that the Cosmic Lithium Problem will be brought to a conclusion. Many, but not all, of the community suspect that lithium is destroyed during the lives of even the most metal-poor stars during either the pre-main sequence, or main sequence phases of its life. If this is so, then we can only determine the true primordial lithium abundance if we can: (1) develop improved and reliable models of \lis\ astration that allow us to correct for the `missing' lithium; or (2) develop new approaches (such us interstellar absorption) to obtain a completely independent measurement of the primordial lithium abundance, that is not limited by the astration of lithium.

Finally, while it is quite unlikely that any other primordial nuclide will be detected and utilized to test models of the early Universe, the most likely of these are perhaps $^{6}$Li, $^{9}$Be, and $^{10}$B, where there have been several efforts to measure these abundances in the atmospheres of metal-poor stars. For all intents and purposes, it will be essentially impossible to detect these isotopes in a primitive interstellar medium.

\section{Concluding remarks}

\begin{figure}
    \centering
    \includegraphics{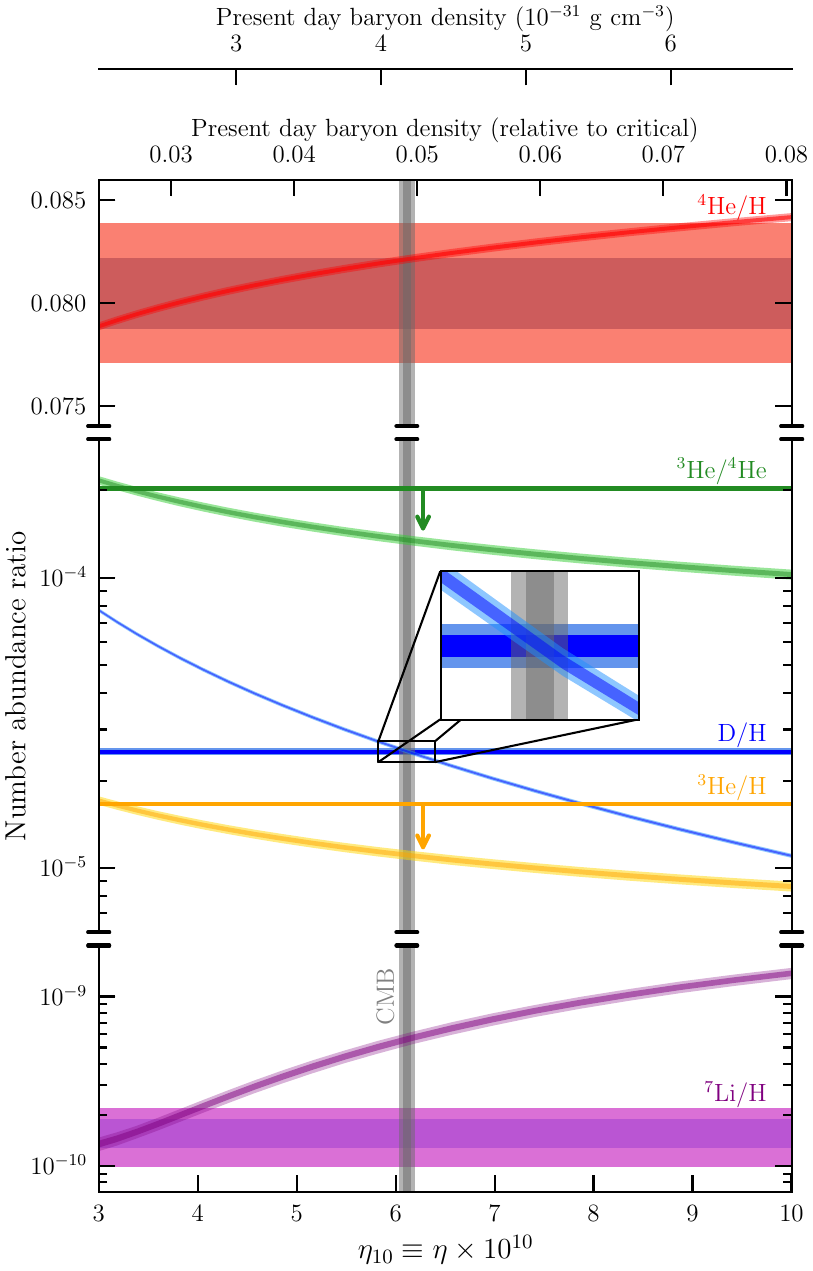}
    \caption{The freeze-out primordial abundances of the dominant nuclides that were produced during BBN are shown for various choices of the baryon-to-photon ratio, $\eta$. The top x-axes indicate the present day density of baryons ($\Omega_{B,0}$) in cgs units (upper axis) and relative to the critical density (lower axis). The horizontal dark and light shaded bands correspond to the $1\sigma$ and $2\sigma$ confidence intervals of current observational estimates of the primordial abundances. $2\sigma$ upper limits on the primordial \het/\hef\ and \het/H abundances are shown with arrows. The current determination of the baryon-to-photon ratio based on the CMB is shown as the vertical grey band (dark and light shades indicate the $1\sigma$ and $2\sigma$ confidence intervals). The coloured curves that depend on $\eta$ are based on Standard Model primordial nucleosynthesis calculations, where the widths of the bands indicate the uncertainties in the calculation. The inset figure shows a zoom-in of the intersection between the CMB, BBN calculations, and a measurement of the primordial D/H value. Note the different y-axis scales that are used to emphasize the sensitivity of the abundances to $\eta$.}
    \label{fig:abundance_eta}
\end{figure}

In this chapter, we have covered the current status of both the theory and observational techniques relevant to Big Bang Nucleosynthesis. Given the content that is covered by this Chapter, a summary of the current state of BBN is shown in Figure~\ref{fig:abundance_eta}. The horizontal bands show the current observational bounds on the primordial abundances, while the vertical grey band shows the baryon-to-photon ratio measured from the CMB. The diagonal bands show the calculated values of the primordial abundances for several different choices of baryon-to-photon ratio based on detailed calculations of BBN. Both the primordial \hef/H and D/H measures are in near-perfect agreement, and both of the limits on the \het/H and \het/\hef\ ratios are consistent with the Standard Model curves. As discussed in Section~\ref{sec:lithium}, the primary tension with BBN is the Cosmic Lithium Problem, which is the disagreement between measures of the primordial $^{7}$Li/H abundance, the baryon-to-photon ratio, and BBN calculations.

To conclude this chapter, there is something very important that cannot be overstated: It is truly impressive that our understanding of cosmology, astrophysics, particle physics, and nuclear physics have all been combined to develop a wonderful experiment that allows us to study the first few minutes after the Big Bang. Remarkably, measurements of the primordial nuclides that we have obtained so far are in excellent agreement with the Standard Model values --- in some cases, the agreement is better than one percent. The only aspect of this field that is more exciting than the progress that has been made so far, is the future potential of BBN to discover new physics beyond the Standard Model.

\begin{ack}[Acknowledgments]

The author notes that some of the material presented in this Chapter was initially reviewed in `A Cosmology Workbook', and has been significantly updated here.
%RC would like to thank BLAH for comments on a draft version of this chapter.
RC is funded by a Royal Society University Research Fellowship, and acknowledges support from STFC (ST/T000244/1).

\end{ack}

% \seealso{article title article title}

\bibliographystyle{Harvard}
\bibliography{els-article}

\begin{appendix}
\setcounter{secnumdepth}{0}

\section{Further Information}
This section contains a list of useful references for readers that wish to explore Big Bang nucleosynthesis in more detail. This is far from an exhaustive list, but this compilation should provide an overall view of the history and current state of the field. The aim is to provide the reader with some stepping stones into the relevant literature.

\subsection{Review articles}
There are many excellent reviews and books that cover different aspects of Big Bang nucleosynthesis. An incomplete list of reviews include those by \citepFurther{Steigman2007,Iocco2009,Cyburt2016,Fields2020}. \citetFurther{SchrammTurner1998} offer a review that covers some of the earlier developments of BBN. A thorough review of the Cosmic Lithium Problem is covered by \citetFurther{Fields2011}, and there are several review articles that also include discussions about physics beyond the Standard Model \citepFurther{PospelovPradler2010,Steigman2012,GrohsFuller2022}. Finally, there are many cosmology textbooks that include one chapter on Big Bang nucleosynthesis, but see the book by  \citetFurther{KolbTurner1990} for a thorough summary, and the more modern textbook by \citetFurther{DodelsonSchmidt2020} is also recommended.

\subsection{BBN theory and calculations}

Several public codes are available to compute the primordial abundances of the light nuclides, including
\texttt{PArthENoPE} \citepFurther{Pisanti2008,Consiglio2018,Gariazzo2022},
\texttt{AlterBBN} \citepFurther{Arbey2012,Arbey2020},
\texttt{PRIMAT} \citepFurther{Pitrou2018,Pitrou2021},
\texttt{PRyMordial} \citepFurther{Burns2024}, and
\texttt{LINX} \citepFurther{Giovanetti2024}. These software packages are based on different programming languages, implement different rates, and the choice of how to implement these rates is different. Overall, there is good general agreement among different codes, with the possible exception of the final predicted abundance of deuterons. The current disagreement between different codes is largely due to the implementation of the $d+d$ rates (Equations~\ref{eqn:ddn} and \ref{eqn:ddp}). For a further discussion about this current tension, see \citetFurther{Pitrou2021}, \citetFurther{Pisanti2021} and \citetFurther{Yeh2021}. The input parameters and output abundances of BBN calculations are often compared to the cosmological parameters that are inferred from the cosmic microwave background. A discussion about Big Bang nucleosynthesis and the cosmic microwave background is provided in the paper by the \citetFurther{Planck2020}.

\subsection{Measures of the primordial deuterium abundance}

There are just a small handful of detections of deuterium absorption lines in near-pristine gas clouds. The following list includes the reported discoveries of such systems, together with studies that have performed a reanalysis of these systems \citepFurther{BurlesTytler1998a,BurlesTytler1998b,PettiniBowen2001,OMeara2001,Kirkman2003,Crighton2004,OMeara2006,Pettini2008,Fumagalli2011,PettiniCooke2012,Noterdaeme2012,Cooke2014,RiemerSorensen2015,Cooke2016,Balashev2016,RiemerSorensen2017,Zavarygin2018,Cooke2018,Guarneri2024,Kislitsyn2024}. Deuterium absorption has also been detected along multiple sightlines that probe the Milky Way interstellar medium \citepFurther{Linsky2006}.

\subsection{Measures of the primordial $^{4}$He abundance}

There are many excellent papers on the measurement of the primordial $^{4}$He abundance. The following list aims to provide readers with a starting point to understand the different approaches that have recently been developed by different groups to pin down this fundamental measurement \citepFurther{Izotov2014,Peimbert2016,Fernandez2019,Valerdi2019,Hsyu2020,Kurichin2021,Aver2021,Matsumoto2022}.

\subsection{Measures of the primordial $^{3}$He abundance}

Since measurements of $^{3}$He are so rare, it is worthwhile mentioning the range of environments where $^{3}$He has been detected. There are several approaches to measure the $^{3}$He abundance in the Solar System, including Earth’s mantle \citepFurther{Peron2018}, Jupiter \citepFurther{Mahaffy1998}, meteorites \citepFurther{Busemann2000,Krietsch2021}, solar wind particles \citepFurther{Heber2012}, and the Local Interstellar Cloud \citepFurther{Busemann2006}. $^{3}$He has also been detected in some planetary nebulae \citepFurther{BaniaBalser2021}, as well as in \HII\ regions of the Milky Way by the hyperfine transition \citepFurther{Bania2002,BalserBania2018} and optical/near-infrared transitions \citepFurther{Cooke2022}. The theory behind a possible measure of $^{3}$He in the intergalactic medium has been explored \citepFurther{McQuinnSwitzer2009,Takeuchi2014,Khullar2020}, and this may still be quite challenging, even with forthcoming facilities such as the Square Kilometre Array.

\subsection{Measures of the primordial $^{7}$Li abundance}

There is a long history of primordial $^{7}$Li abundance measurements; an excellent historical summary of these measurements has recently been discussed by \citetFurther{Norris2023}. Overall, the Spite plateau has remained impressively constant for decades, with most authors agreeing upon the value. Several early works on the most likely explanation of the Cosmic Lithium Problem include those by \citetFurther{Richard2005,Korn2006,Piau2006,Lind2009}. Discussions about the lithium meltdown in extremely metal-poor stars can be found in the following papers \citepFurther{Sbordone2010,Bonifacio2012,Mucciarelli2022}. As an alternative to the stellar abundance measurements, gas-phase lithium abundance determinations have also been reported recently in metal-poor environments by \citetFurther{Howk2012} and \citetFurther{Molaro2024}. Finally, it is noteworthy that some observations of metal-poor stars revealed a possible $^{6}$Li plateau in significant disagreement with the Standard Model, but subsequent improved modelling revealed that the $^{6}$Li detections were spurious \citepFurther{Lind2013}.

\section{Relevant Websites}
For readers interested in exploring more of the topics covered in this chapter, please see the following list of relevant websites:

\begin{itemize}
    \item \texttt{PArthENoPE} BBN calculation --- \url{https://parthenope.na.infn.it/}
    \item \texttt{AlterBBN} BBN calculation --- \url{https://alterbbn.hepforge.org/}
    \item \texttt{PRIMAT} BBN calculation --- \url{https://www2.iap.fr/users/pitrou/primat.htm}
    \item \texttt{PRyMordial} BBN calculation --- \url{https://github.com/vallima/PRyMordial}
    \item \texttt{LINX} BBN calculation --- \url{https://github.com/cgiovanetti/LINX}
    \item Stellar Abundances for Galactic Archaeology (SAGA) Database (see the left panel of Figure~\ref{fig:heliummetals}) --- \url{http://sagadatabase.jp/}
\end{itemize}

\end{appendix}

\bibliographystyleFurther{Harvard}
\bibliographyFurther{further}

\end{document}